\documentclass[twocolumn]{aastex61}

\usepackage{graphicx}
\usepackage{amssymb}
\usepackage{amsmath}
\usepackage{color}
\usepackage{float}
\usepackage{enumerate}
\usepackage{natbib}
\newcommand{\s}{$\sim$}
\newcommand{\dg}{$^\circ$}
\newcommand{\as}{\arcsec{}}

\begin{document}
\title{The circumstellar disk HD\,169142: gas, dust and planets acting in concert?\footnote{Based on observations collected at the European Organisation for Astronomical Research in the Southern Hemisphere under ESO programme 095.C-0273.}}

\correspondingauthor{Adriana~Pohl}
\affiliation{Max-Planck-Institute for Astronomy, K\"onigstuhl 17, D-69117 Heidelberg, Germany}
\affiliation{Heidelberg University, Institute of Theoretical Astrophysics, Albert-Ueberle-Str. 2, D-69120 Heidelberg, Germany}
\email{pohl@mpia.de}

\author{A.~Pohl}
\affiliation{Max-Planck-Institute for Astronomy, K\"onigstuhl 17, D-69117 Heidelberg, Germany}
\affiliation{Heidelberg University, Institute of Theoretical Astrophysics, Albert-Ueberle-Str. 2, D-69120 Heidelberg, Germany}
\author{M.~Benisty}
\affiliation{Unidad Mixta Internacional Franco-Chilena de Astronom\'{i}a, CNRS/INSU UMI 3386 and Departamento de Astronom\'{i}a, Universidad de Chile, Casilla 36-D, Santiago, Chile}
\affiliation{Univ. Grenoble Alpes, CNRS, IPAG, F-38000 Grenoble, France}
\author{P.~Pinilla}
\affiliation{Department of Astronomy/Steward Observatory, The University of Arizona, 933 North Cherry Avenue, Tucson, AZ 85721, USA}
\author{C.~Ginski}
\affiliation{Leiden Observatory, Leiden University, PO Box 9513, 2300 RA, Leiden, The Netherlands}
\affiliation{Anton Pannekoek Institute for Astronomy, University of Amsterdam, Science Park 904, 1098 XH Amsterdam, The Netherlands}
\author{J.~de~Boer}
\affiliation{Leiden Observatory, Leiden University, PO Box 9513, 2300 RA, Leiden, The Netherlands}
\author{H.~Avenhaus}
\affiliation{Institute for Astronomy, ETH Zurich, Wolfgang-Pauli-Strasse 27, 8093 Zurich, Switzerland}
\author{Th.~Henning}
\affiliation{Max-Planck-Institute for Astronomy, K\"onigstuhl 17, D-69117 Heidelberg, Germany}
\author{A.~Zurlo}
\affiliation{Aix Marseille Universit\'{e}, CNRS, LAM (Laboratoire d'Astrophysique de Marseille) UMR 7326, 13388, Marseille, France}
\affiliation{N\'{u}cleo de Astronom\'{i}a, Facultad de Ingenier\'{i}a, Universidad Diego Portales, Av. Ejercito 441, Santiago, Chile}
\affiliation{Departamento de Astronom\'{i}a, Universidad de Chile, Casilla 36-D, Santiago, Chile}
\author{A.~Boccaletti}
\affiliation{LESIA, Observatoire de Paris, PSL Research University, CNRS, Sorbonne Universit\'{e}s, UPMC Univ. Paris 06, Univ. Paris Diderot, Sorbonne Paris Cit\'{e}, 5 place Jules Janssen, 92195 Meudon, France}
\author{J.-C.~Augereau}
\affiliation{Univ. Grenoble Alpes, CNRS, IPAG, F-38000 Grenoble, France}
\author{T.~Birnstiel}
\affiliation{University Observatory, Faculty of Physics, Ludwig-Maximilians-Universit\"{a}t M\"{u}nchen, Scheinerstr. 1, 81679 M\"{u}nchen, Germany} 
\author{C.~Dominik}
\affiliation{Anton Pannekoek Institute for Astronomy, University of Amsterdam, Science Park 904, 1098 XH Amsterdam, The Netherlands}
\author{S.~Facchini}
\affiliation{Max-Planck-Institut f\"ur Extraterrestrische Physik, Giessenbachstrasse 1, 85748 Garching, Germany}
\author{D.~Fedele}
\affiliation{INAF-Osservatorio Astrofisico di Arcetri, L.go E. Fermi 5, I-50125 Firenze, Italy}
\author{M.~Janson}
\affiliation{Max-Planck-Institute for Astronomy, K\"onigstuhl 17, D-69117 Heidelberg, Germany}
\affiliation{Department of Astronomy, Stockholm University, AlbaNova University Center, 10691 Stockholm, Sweden}
\author{M.~Keppler}
\affiliation{Max-Planck-Institute for Astronomy, K\"onigstuhl 17, D-69117 Heidelberg, Germany}
\author{Q.~Kral}
\affiliation{Institute of Astronomy, University of Cambridge, Madingley Road, Cambridge CB3 0HA, UK}
\author{M.~Langlois}
\affiliation{CRAL, UMR 5574, CNRS, Universit\'{e} Lyon 1, 9 avenue Charles Andr\'{e}, 69561 Saint Genis Laval Cedex, France}
\affiliation{Aix Marseille Universit\'{e}, CNRS, LAM (Laboratoire d'Astrophysique de Marseille) UMR 7326, 13388, Marseille, France}
\author{R.~Ligi}
\affiliation{Aix Marseille Universit\'{e}, CNRS, LAM (Laboratoire d'Astrophysique de Marseille) UMR 7326, 13388, Marseille, France}
\author{A.-L.~Maire}
\affiliation{Max-Planck-Institute for Astronomy, K\"onigstuhl 17, D-69117 Heidelberg, Germany}
\author{F.~M\'{e}nard}
\affiliation{Univ. Grenoble Alpes, CNRS, IPAG, F-38000 Grenoble, France}
\author{M.~Meyer}
\affiliation{Institute for Astronomy, ETH Zurich, Wolfgang-Pauli-Strasse 27, 8093 Zurich, Switzerland}
\author{C.~Pinte}
\affiliation{Univ. Grenoble Alpes, CNRS, IPAG, F-38000 Grenoble, France}
\author{S.~P.~Quanz}
\affiliation{Institute for Astronomy, ETH Zurich, Wolfgang-Pauli-Strasse 27, 8093 Zurich, Switzerland}
\author{J.-F.~Sauvage}
\affiliation{ONERA, Optics Department, BP 72, F-92322 Chatillon, France}
\author{\'{E}.~Sezestre}
\affiliation{Univ. Grenoble Alpes, CNRS, IPAG, F-38000 Grenoble, France}
\author{T.~Stolker}
\affiliation{Anton Pannekoek Institute for Astronomy, University of Amsterdam, Science Park 904, 1098 XH Amsterdam, The Netherlands}
\author{J.~Szul\'{a}gyi}
\affiliation{Institute for Astronomy, ETH Zurich, Wolfgang-Pauli-Strasse 27, 8093 Zurich, Switzerland}
\author{R.~van~Boekel}
\affiliation{Max-Planck-Institute for Astronomy, K\"onigstuhl 17, D-69117 Heidelberg, Germany}
\author{G.~van~der~Plas}
\affiliation{Univ. Grenoble Alpes, CNRS, IPAG, F-38000 Grenoble, France}
\author{A.~Baruffolo}
\affiliation{INAF-Osservatorio Astronomico di Padova, Vicolo dell'Osservatorio 5, I-35122 Padova, Italy}
\author{P.~Baudoz}
\affiliation{LESIA, Observatoire de Paris, PSL Research University, CNRS, Sorbonne Universit\'{e}s, UPMC Univ. Paris 06, Univ. Paris Diderot, Sorbonne Paris Cit\'{e}, 5 place Jules Janssen, 92195 Meudon, France}
\author{D.~Le~Mignant}
\affiliation{Aix Marseille Universit\'{e}, CNRS, LAM (Laboratoire d'Astrophysique de Marseille) UMR 7326, 13388, Marseille, France}
\author{D.~Maurel}
\affiliation{Univ. Grenoble Alpes, CNRS, IPAG, F-38000 Grenoble, France}
\author{J.~Ramos}
\affiliation{Max-Planck-Institute for Astronomy, K\"onigstuhl 17, D-69117 Heidelberg, Germany}
\author{L.~Weber}
\affiliation{Geneva Observatory, University of Geneva, Chemin des Maillettes 51, 1290 Versoix, Switzerland}

\shorttitle{The circumstellar disk HD\,169142: gas, dust and planets acting in concert?}
\shortauthors{Pohl et al.}

\begin{abstract}
HD\,169142 is an excellent target to investigate signs of planet-disk interaction due to the previous evidence of gap structures. We performed $J$-band (\s1.2\,$\mu$m) polarized intensity imaging of HD\,169142 with VLT/SPHERE. We observe polarized scattered light down to 0\farcs16 ($\sim$19\,au) and find an inner gap with a significantly reduced scattered light flux. We confirm the previously detected double ring structure peaking at 0\farcs18 ($\sim$21\,au) and 0\farcs56 ($\sim$66\,au), and marginally detect a faint third gap at 0\farcs70-0\farcs73 ($\sim$82-85\,au). We explore dust evolution models in a disk perturbed by two giant planets, as well as models with a parameterized dust size distribution. The dust evolution model is able to reproduce the ring locations and gap widths in polarized intensity, but fails to reproduce their depths. It, however, gives a good match with the ALMA dust continuum image at 1.3\,mm. Models with a parameterized dust size distribution better reproduce the gap depth in scattered light, suggesting that dust filtration at the outer edges of the gaps is less effective. The pile-up of millimeter grains in a dust trap and the continuous distribution of small grains throughout the gap likely require a more efficient dust fragmentation and dust diffusion in the dust trap. Alternatively, turbulence or charging effects might lead to a reservoir of small grains at the surface layer that is not affected by the dust growth and fragmentation cycle dominating the dense disk midplane. The exploration of models shows that extracting planet properties such as mass from observed gap profiles is highly degenerate.
\end{abstract}


\section{Introduction}
\label{sec:intro}
About two decades ago, our own Solar System was the only available laboratory to test models of planet formation. Today, we know that planetary systems are common around other stars, and that their architectures are very diverse. The initial conditions and evolution of protoplanetary disks, where planets form, must have a direct influence on most fundamental properties of their planetary systems \citep{mordasini2012,mordasini2016}. It is therefore essential to improve our knowledge of the structure of protoplanetary disks by observing and studying them at high spatial scales, and with various tracers that enable to characterize different disk regions. This, indirectly, can constrain the physical processes that influence the disk evolution (e.g., gap opening by a planet, dust growth \& settling, photo-evaporation). Even with the advent of a new generation of extreme adaptive optics instruments, the detection of forming planets within their host disks is still challenging, but one can look for indirect signatures of planet formation, such as the imprints that it leaves on the disk.

In recent years, high-resolution images of protoplanetary disks have been published, both in scattered light that trace the (sub-)micron-sized dust particles in the upper disk layers, and in the sub-millimeter (mm) regime that trace larger dust grains (mm and centimeter-sized (cm) grains), while the bulk mass is not directly observable. While for a long time protoplanetary disks were thought to be smooth and continuous, a variety of small scale features are now frequently detected in these images, and seem rather common, if not ubiquitous. Large cavities ($\sim$few tens of au) are detected in a number of objects \citep[e.g.,][]{williams2011}, the transition disks, that often have spectral energy distributions (SEDs) with a clear dip in the mid-infrared (MIR), indicating a lack of dust in the inner regions \citep[][]{strom1989}. Smaller cavities and gaps in the inner astronomical units (au) are also present \citep{menu2015} but cannot be easily directly imaged, nor do they leave a clear imprint on the SED. Multiple-arm spiral features were observed, mostly in scattered light \citep{muto2012,grady2013,garufi2013,avenhaus2014,benisty2015,stolker2016,benisty2017}, but also more recently, in the sub-mm wavelength range (CO gas lines: \citealt{christiaens2014,tang2017}; continuum emission: \citealt{perez2016}). These spiral arms have large opening angles, and their origin is still not fully understood. The presence of one or multiple rings and gaps in disks seems quite a common feature, they are found in both young \citep[e.g. HL Tau,][]{alma2015,carrasco2016} and rather old systems \citep[e.g. TW Hya,][]{andrews2016,tsukagoshi2016,rapson2015,vanboekel2017}, and around stars of very different spectral types \citep[e.g.,][]{deboer2016,ginski2016,vanderplas2017}. Various mechanisms have been proposed in the literature, which can be assigned to three main categories: structures caused by fluid dynamics, dust evolution effects, and planet-disk perturbations. More precisely, these possibilities include zonal flows from magneto-rotational instability \citep[e.g.,][]{simon2014,bethune2016}, gap/bump structures in the surface density close to the dead-zone outer edge \citep[e.g.,][]{flock2015,ruge2016,pinilla2016}, efficient particle growth at condensation fronts near ice lines or a depletion of solid material between ice lines \citep[][]{zhang2015,stammler2017,pinilla2017}, aggregate sintering zones \citep{okuzumi2016}, secular gravitational instabilities \citep{youdin2011,takahashi2014}, or planet-disk interactions \citep[e.g.][]{zhu2011,zhu2012,dong2015b,dong2016,rosotti2016}. Finally, dips or dark regions can be interpreted as shadows by inner disk material \citep[e.g.,]{marino2015,pinilla2015a, stolker2016,canovas2017}.

We focus on a multiple-ring system in this study, more specifically, the $\sim$6$^{+6}_{-3}$\,Myr old Herbig A5/A8 star HD\,169142 \citep{dunkin1997,grady2007}, located at  a distance of d=117$\pm$4\,pc\footnote{Note that we are using the revised value by GAIA while most of the papers in the literature use d=145\,pc.} \citep{gaia}. With this new distance of 117\,pc, the star is intrinsically less luminous by a factor of $\sim$0.65. The age estimate by \citet{grady2007} is based on Hertzsprung-Russell (HR) placement of the companion 2MASS~18242929--2946559. Moving this star down in the HR diagram \citep[Fig.~9 in][]{grady2007} leads to a revised age estimate of $\sim$10 Myr. Its SED shows a strong infrared excess indicating a young gas-rich disk, with many emission line features \citep{rivieremarichalar2016,kama2016,seok2017}, and a clear dip of emission in the infrared regime \citep{grady2007,meeus2010}, qualifying it as a transition disk. The near-infrared (NIR) flux indicates the presence of hot dust close to the sublimation radius, resolved by NIR interferometric observations \citep[][]{lazareff2017}. HD\,169142 still experiences gas accretion onto the star, with estimates of the mass accretion rate varying between 0.7 and 2.7 $\times 10^{-9}\,M_{\odot}\,\mathrm{yr}^{-1}$ \citep{grady2007,wagner2015}. \citet{garufi2017} note that HD\,169142 has a reduced NIR excess compared to continuous Herbig disks or those hosting spirals. The NIR and MIR fluxes were also found to vary by up to $\sim$45\% over a temporal baseline of ten years, indicating strong variability in the innermost regions \citep{wagner2015}.

The outer disk has a low inclination (i=13\dg, PA=5\dg) as derived by CO mm observations \citep{raman2006,panic2008}, and confirmed with high-contrast imaging in the near-infrared \citep{quanz2013, momose2015, monnier2017}. These images show, from small to larger separations from the star, a wide inner cavity, a bright (unresolved) ring, a second wide gap and an outer disk that extends up to 1.7\as. The inner cavity appears devoid of small dust grains, while the second gap is not. Observations with ALMA at 1.3\,mm, obtained with a resolution of 0\farcs28$\times$0\farcs18, also show two rings (0\farcs17--0\farcs28 and 0\farcs48--0\farcs64) and a gap between them \citep{fedele2017}. The mm continuum extends up to 0\farcs64 while the gas extends up to twice as far. The channel maps of the 2--1 line transition of the three CO isotopologues reveal the presence of gas inside the dust gaps. Model fitting provides a drop in the gas surface density by a factor of 30--40. The two rings are also detected in Very Large Array observations at longer wavelengths \citep[7\,mm;][]{osorio2014,macias2017} and the azimuthally averaged radial intensity profiles indicate the marginal detection of a new gap at $\sim$0\farcs7, very close to the CO ice line \citep{macias2017}. In addition to the disk features, a candidate massive companion was proposed, slightly inside the inner ring, at a separation of \s0\farcs11 and \s0\farcs16, respectively \citep{biller2014,reggiani2014}. \citet{osorio2014} report the detection of a compact 7\,mm emission source with VLA external to the inner ring. The detection of point-like structures in the context of potential planetary companions is discussed further in \citet{ligi2017}.

In this paper, we report new polarized differential images of HD\,169142 obtained in the $J$-band with the SPHERE instrument \citep[Spectro-Polarimeter High-contrast Exoplanet REsearch,][]{beuzit2008} at the Very Large Telescope (VLT), complemented with ALMA continuum data from \citet{fedele2017}. We investigate whether the observed rings can be explained by trapping of dust particles as a consequence of the presence of two planets. This study on HD\,169142 serves as a prototype, with which it is demonstrated that multi-wavelength observations are needed to constrain the dust size distribution and physical mechanisms at work in the disk. The paper is organized as follows. Section~2 describes the observations and the data processing. Section~3 reports on the detected disk features, Sect.~4 provides a physical disk model, and in Sect.~5 we discuss our findings. 

\section{Observations and data reduction}
\label{sec:obs}

\begin{figure*}
	\centering
	\begin{tabular}{cc}
        \includegraphics[width=\columnwidth]{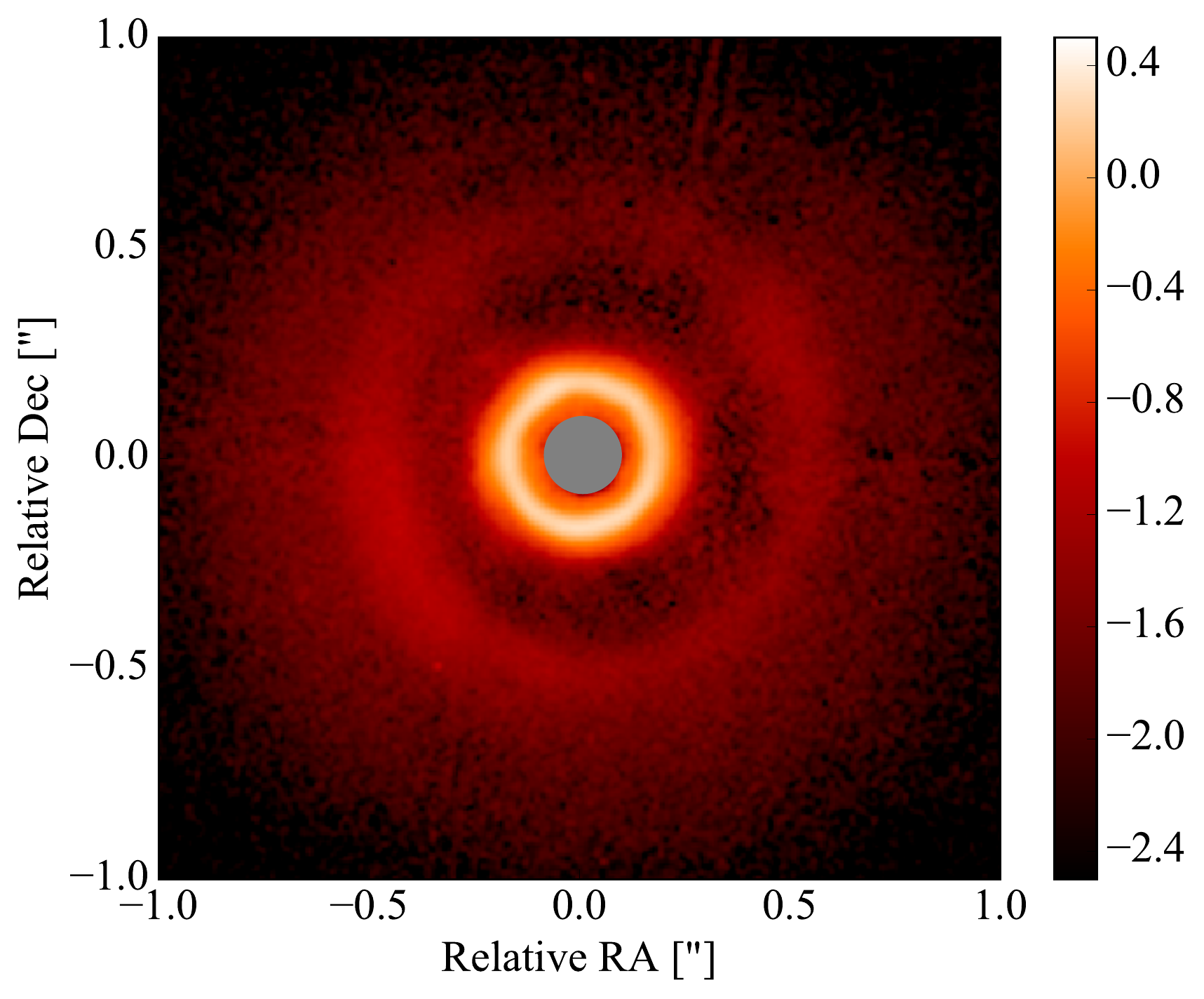} & 
          \includegraphics[width=\columnwidth]{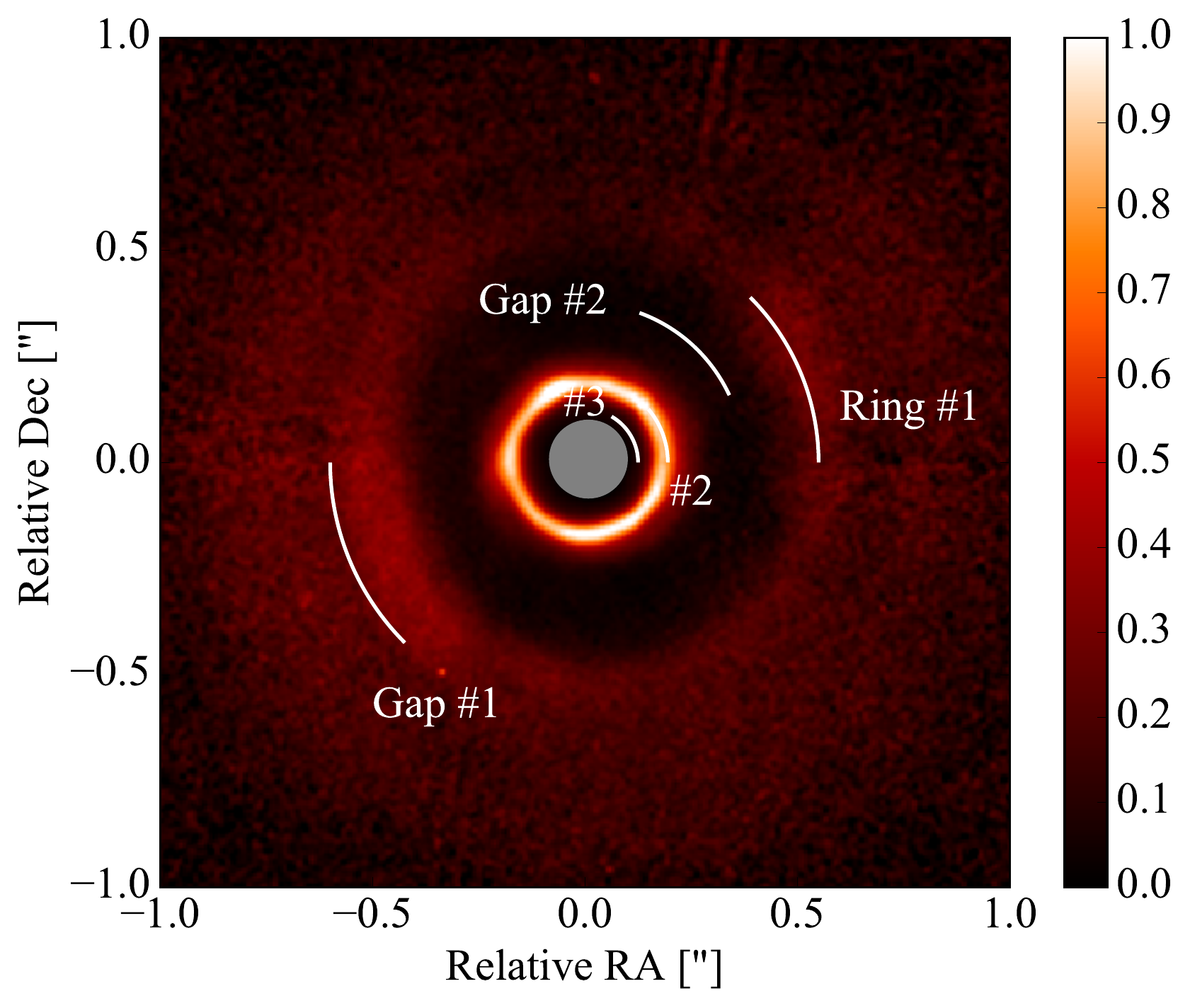} 	
	\end{tabular}
	\caption{\textit{Left:} $J$-band azimuthally polarized intensity image $Q_{\phi}$ in logarithmic scale for better visualization. \textit{Right:} $Q_{\phi} \times r^2$ in linear scale with annotations for the gap and ring structures. Each image pixel is multiplied with the square of its distance to the star, $r^{2}$,  to compensate for the stellar illumination drop-off with radius. All flux scales are normalized to half of the brightest pixel along the inner ring. The region masked by the coronagraph is indicated by the gray circle. North is up, East points towards left.}
	\label{fig:images}
\end{figure*}

\begin{figure*}
	\centering
		\begin{tabular}{cc}
        \includegraphics[width=\columnwidth]{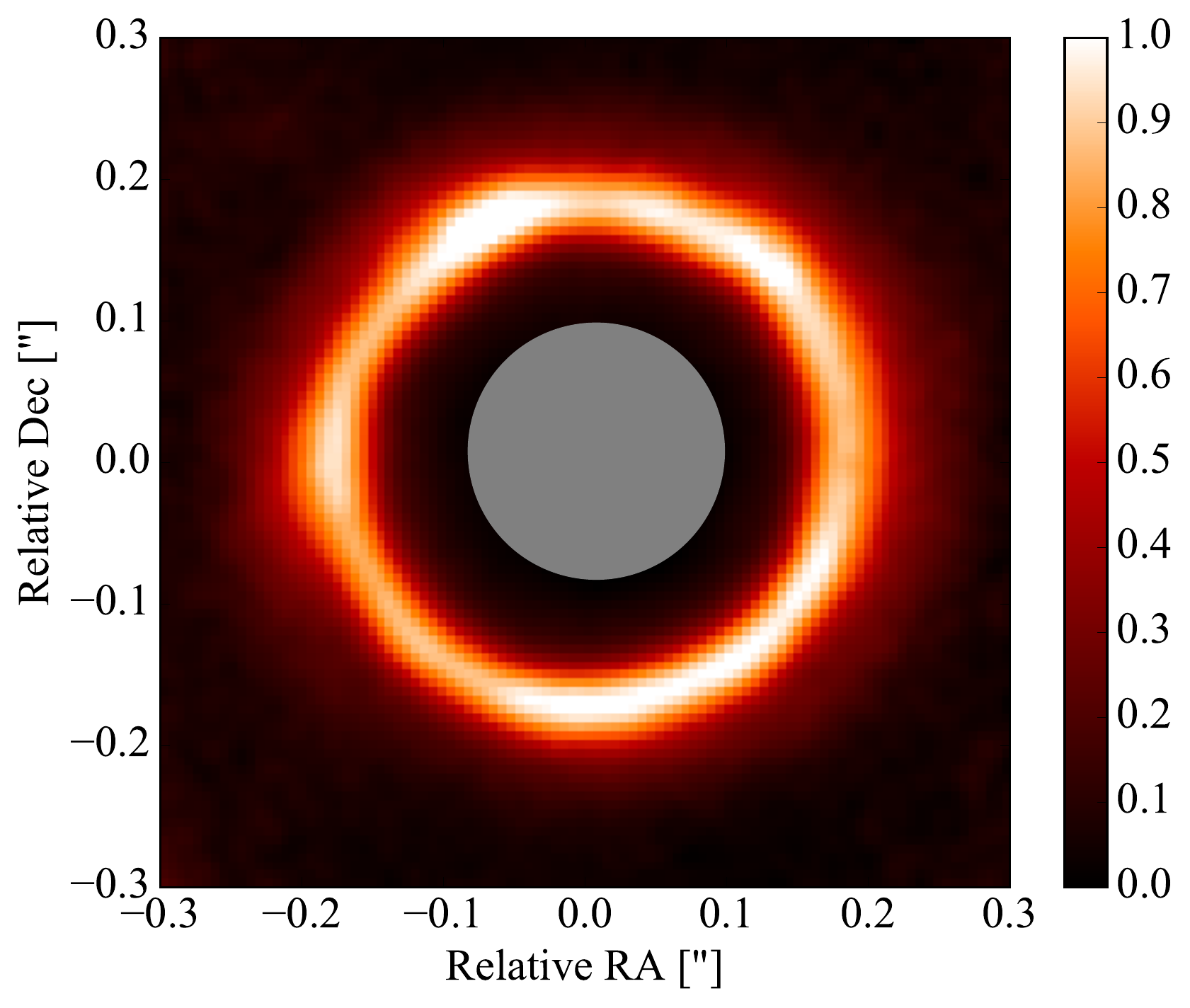}  & 
         \includegraphics[width=\columnwidth]{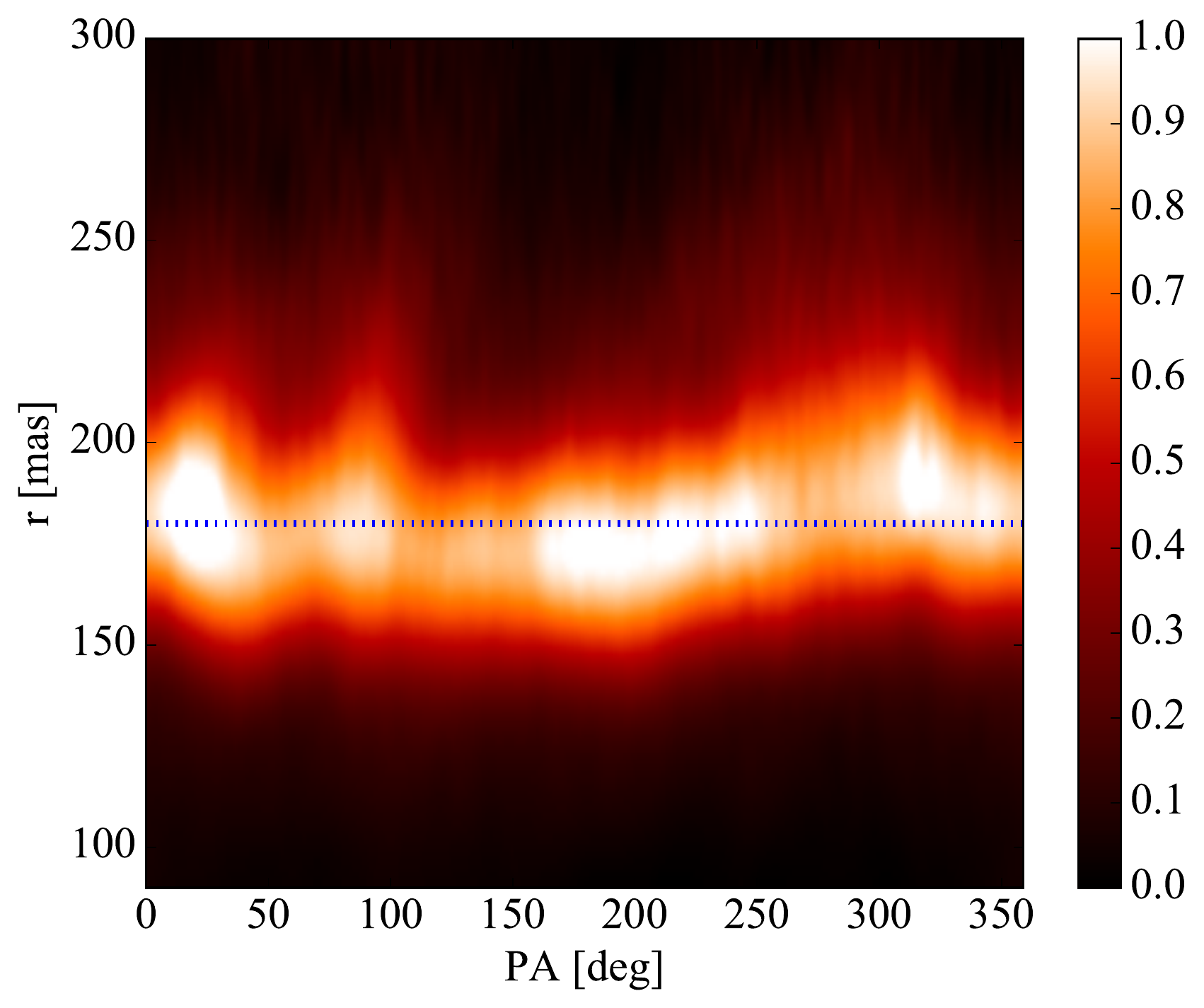}  
	\end{tabular}
	\caption{\textit{Left:} zoom-in on the central 0\farcs3 of the $J$-band $Q_{\phi} \times r^2$ image. \textit{Right:} polar map of the $Q_{\phi} \times r^2$ image. The flux scales are normalized to half of the brightest pixel along the ring. The horizontal dashed line indicates a radius of 0\farcs18.}
	\label{fig:zoom}
\end{figure*}

The observations were obtained at the Very Large Telescope at Cerro Paranal, Chile, on 2015 May 02 with the SPHERE instrument. SPHERE is equipped with an extreme adaptive-optics (AO) system \citep{fusco2006,petit2014,sauvage2014} that feeds three science channels allowing for high-angular resolution and high-contrast imaging, spectroimaging, and/or polarimetry at visible and near-infrared wavelengths. The observations were obtained through the Guaranteed Time program. HD\,169142 was observed in the $J$-band filter ($\lambda_0$=1.258, $\Delta\lambda$=0.197\,$\mu$m) using the polarimetric imaging mode of the infrared dual-band imager and spectrograph \citep[IRDIS;][]{dohlen2008,langlois2014}, with a plate scale of 12.25\,mas per pixel \citep[][]{maire2016}, and a 145\,mas-diameter coronagraphic focal mask (N\_ALC\_YJ\_S, inner working angle (IWA) of 0\farcs08, \citealt{boccaletti2008}). HD\,169142 was observed for $\sim$53 minutes on-source under moderate AO conditions (seeing of 0\farcs9). The analysis of the reference point spread function (PSF) that is estimated from a non-coronagraphic total intensity measurement shows that the observations reach a 33.8\,mas $\times$ 40.8\,mas resolution (FWHM along the x and y directions) and a Strehl ratio of 56\%.

We observed HD\,169142 using the polarimetric differential imaging technique \citep[PDI; e.g.][]{kuhn2001,apai2004} that measures the linear polarization of the light scattered by dust grains in the disk, and enables one to efficiently remove the unpolarized contribution, including the one from the star. This allows to image, with high contrast, the polarized signal from the disk. In this mode, the instrument splits the beam into two orthogonal polarization states. The control of the polarization orientation is performed with a half-wave plate (HWP) that was set to four positions shifted by 22.5$^\circ$ in order to construct a set of linear Stokes images. We reduce the data according to the double difference method \citep{kuhn2001}, and obtain the Stokes parameters $Q$ and $U$. If we assume that there is only one scattering event for each photon, the scattered light from a circumstellar disk seen at low inclination angle is expected to be linearly polarized in the azimuthal direction. It is therefore convenient to describe the polarization vector field in polar coordinates \citep{schmid2006,avenhaus2014}. We therefore define the polar-coordinate Stokes parameters $Q_{\phi}$, $U_{\phi}$ as:
\begin{equation}
Q_{\phi} = +Q \cos(2\phi) + U \sin(2\phi)
\end{equation}
\begin{equation}
U_{\phi} = -Q \sin(2\phi) + U \cos(2\phi)\,, 
\end{equation}
\noindent with $\phi$, the position angle of the location of interest (x, y) with respect to the star location. In this coordinate system, the azimuthally polarized flux appears as a positive signal in the $Q_{\phi}$ image, whereas the $U_{\phi}$ image remains free of disk signal and can be used as an estimate of the residual noise in the $Q_{\phi}$ image \citep{schmid2006}. This is only valid for disks with face-on geometry since multiple scattering effects in inclined disks can cause a considerable physical signal in $U_{\phi}$ \citep[e.g., T\,Cha:][]{pohl2017}. The correction for instrumental polarization is done using a $U_{\phi}$ minimization, by subtracting scaled versions of the total intensity frame from the Stokes Q and U frames. The final data images were corrected for the true North \citep[by rotating them by 1.775$^\circ$ in the counterclockwise direction,][]{maire2016}. We do not attempt to perform an absolute flux calibration of our images due to the inherent problems with measuring flux in PDI images.

\section{Polarized intensity images}
\label{sec:pi}

\begin{figure*}
	\centering
	\begin{tabular}{cc}
	\includegraphics[width=\columnwidth]{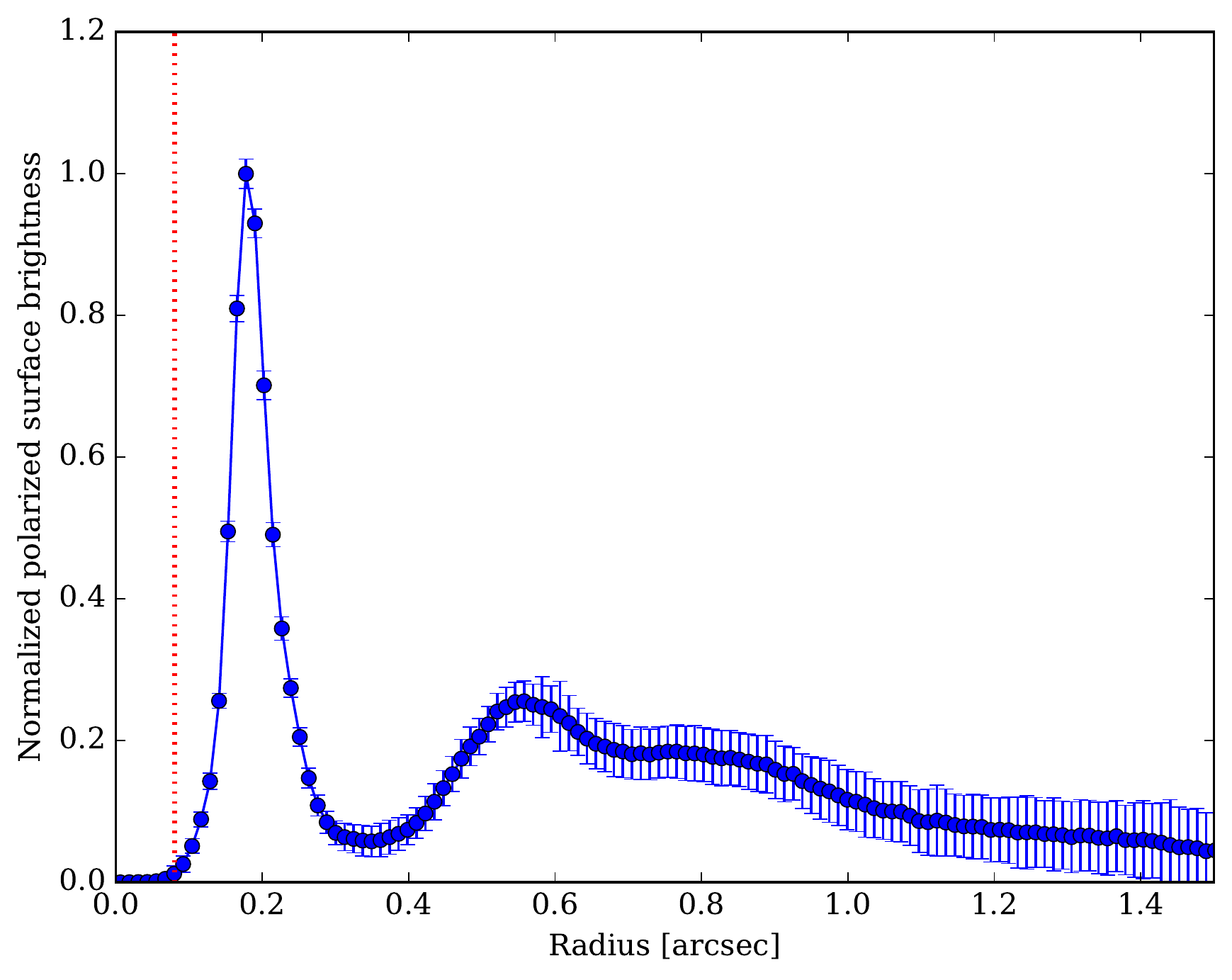} & 
	\includegraphics[width=\columnwidth]{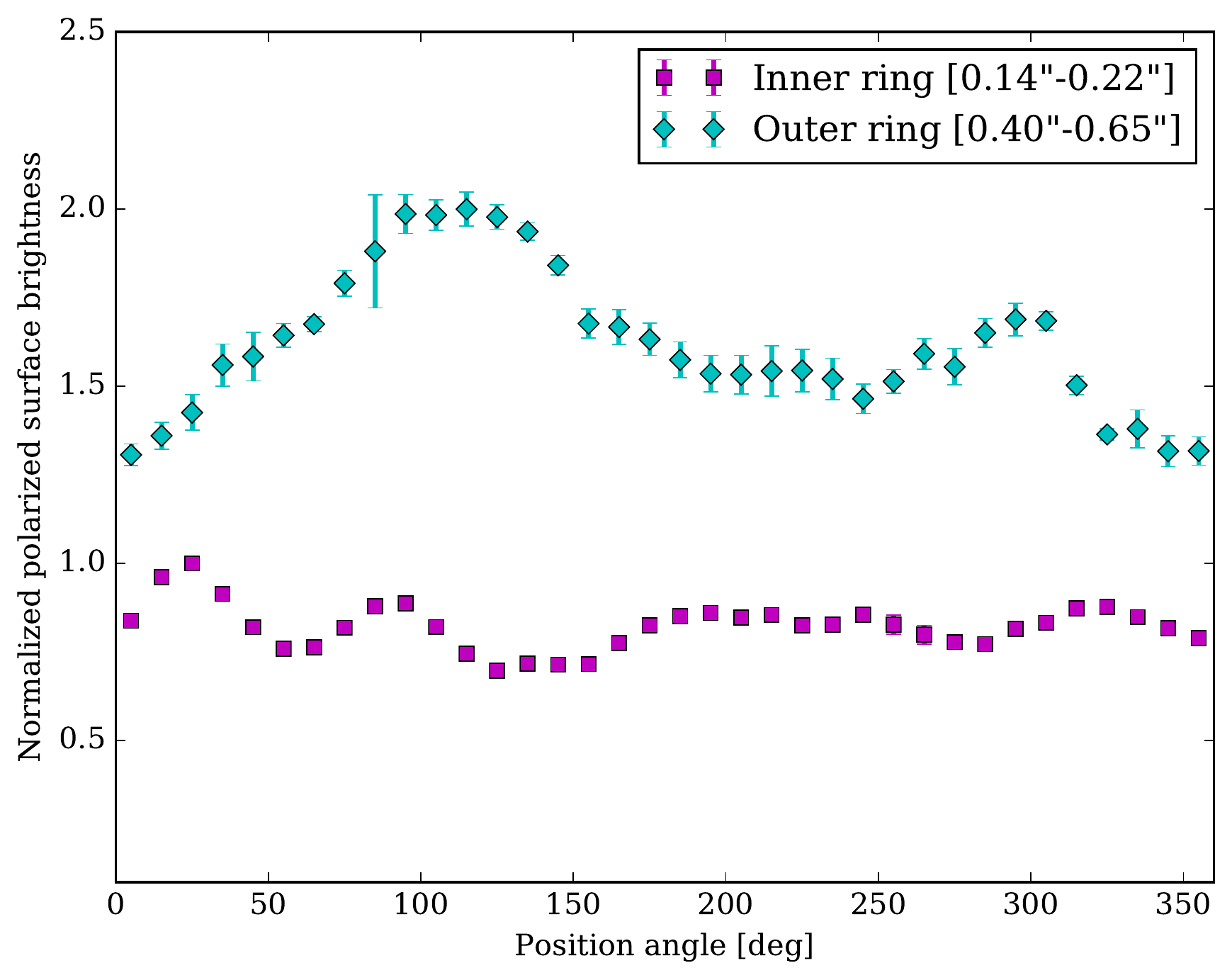}  
	\end{tabular}
	\caption{Normalized radial (\textit{left}) and azimuthal intensity profiles (\textit{right}) obtained after deprojection of the $r^{2}$-scaled $J$-band $Q_{\phi}$ image. The radial cut is obtained after azimuthally averaging and normalized to the maximum brightness of the inner ring. The red vertical dotted line is the limit of our IWA.
	The normalized azimuthal cuts are obtained after averaging radially between 0\farcs14 and 0\farcs22 (inner ring, purple squares), and between 0\farcs4 and 0\farcs65 (outer ring, green diamonds). The green curve is shifted vertically for clarity. The plotted error bars are the standard deviation in each bin in the $U_{\phi}$ image on a pixel basis.}
\label{fig:cut}
\end{figure*}

\begin{figure}
	\centering
	\includegraphics[width=\columnwidth]{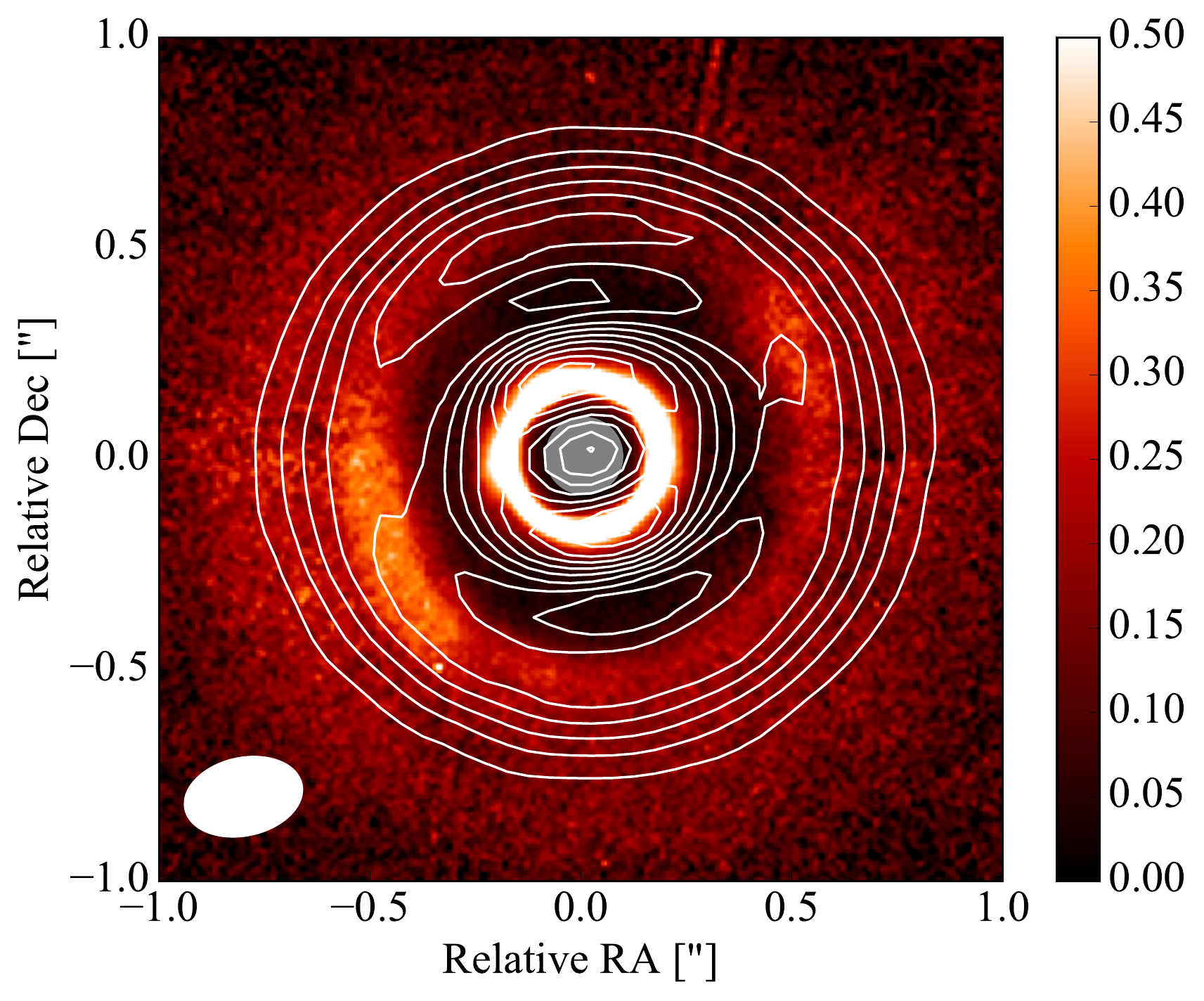}  
	\caption{SPHERE/IRDIS $r^{2}$-scaled $J$-band $Q_{\phi}$ image overlaid with contours of ALMA 1.3\,mm continuum image \citep[from][]{fedele2017}. The white ellipse in the bottom left corner shows the ALMA beam with a size of 0\farcs28$\times$0\farcs18. The $Q_{\phi}$ image is normalized in the same way as Fig.~\ref{fig:images}, right, but the color scale is chosen such that the structures in the outer disk are enhanced and the inner ring is saturated.} 
	\label{fig:overlay}
\end{figure}

Figures \ref{fig:images} and \ref{app:uphi} show the polarized scattered light images $Q_{\phi}$ and $U_{\phi}$, respectively, obtained in the $J$-band. The $U_{\phi}$ image contains very low signal, suggesting that the assumption of single scattering is valid \citep[c.f.][]{canovas2015}. Figure~\ref{fig:images} is similar to previously published scattered light images of HD\,169142 in particular those of \citet{momose2015} and \citet{monnier2017}, but it brings the highest signal-to-noise ratio view of the inner ring. It shows a number of features. We detect from outside in:

(a) A faint gap (Gap\,\#1) at \s0\farcs70-0\farcs73 (81--85\,au). Beyond this radius, the image shows diffuse scattered light up to \s1.5\as (\s175\,au).  
The marginal detection of this gap can be seen in the normalized, azimuthally averaged radial profile of the surface brightness, obtained after deprojection and azimuthally averaging the image (Fig.~\ref{fig:cut}, left).

(b) A ring (Ring\,\#1) peaking at 0\farcs56 (\s66\,au) with an apparent width of \s0\farcs16 (\s19\,au, at PA\s100\dg). This outer ring also shows some azimuthal brightness variation with a dip in scattered light along PA\s-15\dg to 30\dg. This is also detected in the $H$- and $J$-band images of \citet{momose2015} and \citet{monnier2017}.

(c) A wide off-center gap (Gap\,\#2), which width ranges from 0\farcs13 (\s15\,au) along PA$\sim$100$^\circ$ to 0\farcs24 (\s28\,au) along PA$\sim$ 200\dg. In Fig.~\ref{fig:cut}, left panel, it is also evident that this gap is not empty of scattering material with a lowest value of 1--2\% of the peak value at 0\farcs35 (\s41\,au). We note, however, that it could actually be emptier, with light from the adjacent rings inside and outside being convolved into the gap. An additional polar map of the full image showing the various gap widths against position angle is available in the Appendix (Fig.~\ref{app:polar}). 

(d) A resolved bright and narrow ring, called Ring\,\#2, located at 0\farcs18 ($\sim$21\,au) with an apparent width of 40-50\,mas (\s5-6\,au). Its brightness varies azimuthally by up to $\sim$25\%, as evidenced by the zoom displayed in Fig.~\ref{fig:zoom}, left. The regions at PAs $\sim$23\dg, 90\dg, 200\dg, and 315\dg are brighter than the regions at PAs $\sim$0\dg, 60\dg, 130\dg, and 275\dg. Figure~\ref{fig:zoom}, right, shows the image in polar coordinates, after deprojecting it with the inclination and position angle derived from the observed kinematic pattern and line profiles at mm wavelengths (i$\sim$13\dg, PA$\sim$5\dg, respectively). One can see that the ring does not lie on a perfectly horizontal line (at a radius of 0\farcs18 in the plot). This suggests that the ring is intrinsically asymmetric or could be asymmetrically illuminated due to shadowing by the inner disk. The ring might also have a non-negligible vertical extent, although this is rather unlikely due to the face-on disk configuration. A detailed analysis on the geometry of this inner ring based on optical SPHERE-ZIMPOL \citep[Zurich IMaging POLarimeter][]{thalmann2008,schmid2012} data can be found in Bertrang et al., subm.

(e) A region with a deficit of scattered light, called Gap\,\#3, outside of our IWA (0\farcs08). This inner gap appears devoid of scattered light flux, but there is an unresolved inner disk with accretion as discussed in \citet{grady2007} and \citet{wagner2015}.

Fig.\,\ref{fig:cut}, right, shows the azimuthal cuts along the two rings, after deprojecting the $Q_{\phi}$ image, and radially averaging over their apparent widths (between 0\farcs14 and 0\farcs22, and between 0\farcs40 and 0\farcs65, respectively).  One can see that both the inner and outer rings present clear azimuthal variations. The outer disk appears brighter along PA\s110-120\dg, i.e. close to the minor axis of the disk. To characterize better the rings and Gap\,\#2, we attempt to fit ellipses to the image. We follow the procedure described in detail in \citet{deboer2016} and \citet{ginski2016}, and consider 10$^{6}$ annuli for each feature and find the annulus for which the flux is maximized (for the rings) or minimized (for the gap). To reduce the number of free parameters, we fix the inclination and position angle of the ellipses to the values inferred from sub-millimeter interferometry \citep{panic2008}. Our best fit result is shown in the Appendix, in Tab.\,\ref{tab:ellipsesFixed} and Fig.\,\ref{app:polar}. We give the offset of the ellipses from the star position as well as the size of the major and minor axes for each fitted feature. Our error bars are estimated as the standard deviation of the best 1\% fits (i.e. the 1\% fits with the highest flux in the resulting aperture for the rings). All the offsets that we measure are towards the North-West direction \citep[as in][]{momose2015}, which suggests that the South-East side of the disk is the near side of the disk. However, we note that the direction of the offsets is not exactly along the minor axis which might indicate that these offsets do not only trace geometrical effects and that the disk could be eccentric.

The SPHERE/IRDIS $J$-band image is very similar to the $H$-and $J$-band images obtained by \citet{momose2015} with Subaru/HiCIAO and by \citet{monnier2017} with the Gemini Planet Imager (GPI), three years and one year before our observations, respectively. This suggests that the observed azimuthal asymmetries are not due to shadowing from the innermost disk. Dynamical structures in the inner disk would evolve significantly on timescales of years (cf. discussion in Sect.~\ref{subsec:disc_shadow}). The two rings in our image are approximately co-located with the two rings detected in the ALMA millimeter dust continuum \citep{fedele2017} as shown in Figs.~\ref{fig:overlay} and \ref{fig:rtband6}. More precisely, the peaks of the two rings at mm are slightly further out than in our SPHERE scattered light data ($\sim$28\,au and $\sim$70\,au vs. $\sim$21\,au and $\sim$66\,au), consistent with current dust trapping scenarios. Although Gap\,\#2 does not appear devoid of small dust grains, the ALMA image shows no continuum detection in both the inner (Gap\,\#3) and wide (Gap\,\#2) gaps. This indicates that dust particles, independently of their sizes, are filtered out in the inner gap, but that the filtering mechanism at play in the outer gap affects small and large dust grains differently. The inner ring (Ring\,\#2) is also well detected in VLA 7\,mm and 9\,mm observations \citep{osorio2014,macias2017}, although at a slightly larger radius (\s0\farcs21), compared to \s0\farcs18 in scattered light). Furthermore, \citet{macias2017} also report on the detection of a third gap at \s0\farcs73, consistent with the marginal detection in the SPHERE polarized intensity data.

\section{Disk modeling}
\label{sec:modeling}
We start our models by introducing planet-induced depressions in a uniform disk gas density profile to mimic the position and shape of the observed gaps. We present physical simulations including dust evolution and trapping processes to constrain the disk dust distribution and to investigate whether planet-disk interactions are responsible for the detected sub-structures. We take the approach to fix as many parameter values as possible, and do not attempt a best-fitting procedure. Because of the high parameter degeneracy when physical processes related with dust evolution are involved, we do not explore a large grid of these models. Our concept is complementary to the one presented by \citet{monnier2017}, who show a parameterized model without connecting the gap and ring structures to a planetary origin or dust evolution. In their model, the scale height of an inner and outer disk region, and the density scaling factor for the outer gap are determined via a fitting process.

\subsection{Model set-ups}
\label{subsec:dert_methods}

In our models, we consider two spatially separated planets that are massive enough to open a gap in the gas surface density. The planet cores are assumed to be at fixed orbits and are not allowed to migrate. We note that we constrain the total number of planets to two, although multiple low mass planets in close-by resonances could exist to cause the second, wide gap (Gap\,\#2). The perturbed gas surface density profiles $\Sigma_{\mathrm{g}}$ depend on the planet masses and on the disk viscosity. To derive $\Sigma_{\mathrm{g}}$, we consider the analytical solution of \citet{crida2006}, in which the gravitational and pressure torques are assumed to be zero very close to the planet. For this reason, we implement a correction for the depth of the gap using the empirical relation from \citet{fung2014}. The resulting gas surface density distributions are used as inputs to model the dust evolution considering the dust dynamics, including the processes of coagulation, fragmentation, and erosion of dust particles \citep{birnstiel2010,pinilla2015b}. For the background surface density profile we use an exponentially tapered power law with a power index of 1, and a tapered radius of six times the location of the inner planet. As the inner gap appears relatively devoid of scattering material, and free from larger grains, we consider a planet-to-stellar mass ratio of 2$\times$10$^{-3}$ (3.5\,M$_{\rm jup}$) for the inner planet, such that the gap is deep enough to lead to a filtration of particles of all sizes. For the second gap being filled with small particles, we consider the mass of the outer planet close to the mass estimate obtained in \cite{osorio2014} in this region, and we choose 0.7 and 0.3\,M$_{\rm jup}$, the latter being the minimum mass needed to open a gap in the gas surface density (and hence to have a pressure trap at the outer edge of the gap) under our assumptions. The locations of the planets are chosen according to the current SPHERE observations and are $r_1=14$\,au and $r_2=53$\,au, such that the pressure maxima are close to the observed peaks of the mm emission. The companion masses considered in our simulations are compatible with the detection limits obtained in total intensity with IRDIS and the Integral Field Spectrograph (IFS) of SPHERE. These data will be presented in detail in the F100 SPHERE High-Contrast Imaging Survey for Exoplanets (SHINE) data analysis and detection performances paper (Langlois et al., in prep.). The disk temperature profile is a power law \citep[$\sim r^{-0.5}$, cf.~Eq.~25 in][]{birnstiel2010} such that at 1\,au the temperature is $\sim$230\,K. We assume an $\alpha$-viscosity of 10$^{-3}$ throughout the disk, and note that this choice also influences the planet masses assumed as described above. Furthermore, we consider a disk mass of $5 \times 10^{-3}$\,M$_{\odot}$ which is consistent with the value range reported by \citet{panic2008}, and a disk radial extension from 1 to 300\,au. The initial gas-to-dust ratio is 100 and particles are initially 1\,$\mu$m in size. The model follows the evolution of 180 grain sizes (from 1\,$\mu$m to 2\,m) and calculates the dust density distribution at each radius for time scales from 10$^{4}$ to $5 \times 10^{6}$ years. We do not consider the effect of ice lines on the dust dynamics.

To compute synthetic images, we consider the resulting dust distribution as input to the radiative transfer code \textsc{RADMC-3D} \citep{radmc}. From the vertically integrated dust density distribution, we derive the dust density for each grain size $\Sigma_d(r,a)$. From the temperature profile $T(r)$ used in the dust evolution, the pressure scale height $H_p(r)$ is determined. We take the approach in \citet{pohl2016} and calculate the dust scale height for each grain size $a$ following \cite{birnstiel2010} as

\begin{equation}
H_d(r,a) = H_p(r) \times {\rm min} \left(1, \sqrt{\frac{\alpha}{  {\rm min}({\rm St}, 1/2)(1+ {\rm St}^{2})}}\right), 
\end{equation}

\noindent where $\alpha$ is the turbulent viscosity and \textit{St} is the Stokes number, a dimensionless parameter that indicates how well a dust grain is coupled to the gas. In the Epstein regime, valid for most regions of protoplanetary disks and where the molecular hydrogen mean free path is larger than 4/9 times the grain size, the Stokes number at the midplane can be written as

\begin{equation}
{\rm St} = \frac{ \rho_s a}{ \Sigma_{\mathrm{g}}} \frac{\pi}{2}, 
\end{equation}

\noindent with $\rho_{\mathrm{s}}$ the volume density of the dust grain, typically \s 1.2\,g\,cm$^{-3}$ according to the averaged values of the volume density for silicates. Dust grains with sizes corresponding to \textit{St}\s1 are subject to the strongest gas drag and move fast to the regions of pressure maxima \citep{brauer2008}. From the dust surface density and scale height, we compute the volume density profile for each grain size as

\begin{equation}
\rho(R,\phi,z,a) = \frac{ \Sigma_d(R,a)}{\sqrt{2\pi} H_d(R,a)} \exp\left( - \frac{z^{2}}{2 H_{d}(R,a)^{2}}\right), 
\label{eq:volumedens}
\end{equation}

\begin{figure*}
	\centering
	\centerline{
		\includegraphics[width=\textwidth]{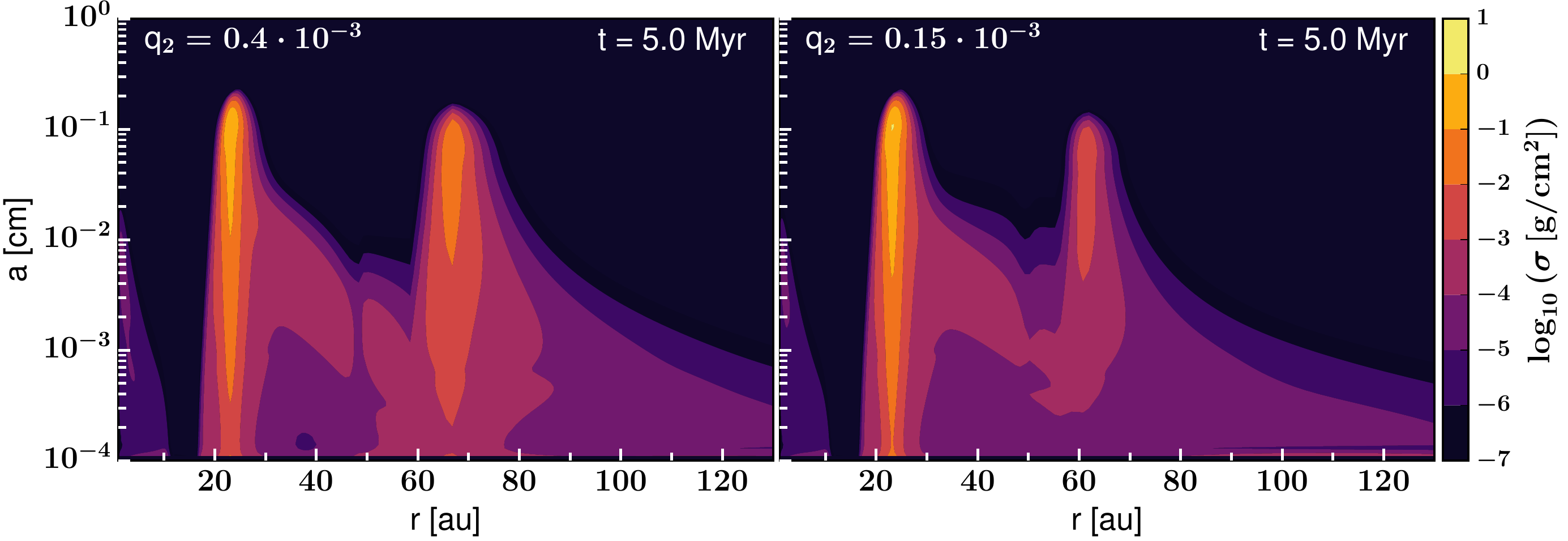}
	}
	\caption{Vertically integrated dust density distribution after 5\,Myr of evolution, when two massive planets (\textit{left:} 3.5 and 0.7\,M$_{\rm jup}$, \textit{right}: 3.5 and 0.3\,M$_{\rm jup}$) are embedded in the disk at 14\,au and 53\,au, respectively.}
	\label{fig:dustdensity1}
\end{figure*}	

\begin{figure*}
	\centering
	\centerline{
		\includegraphics[width=\textwidth]{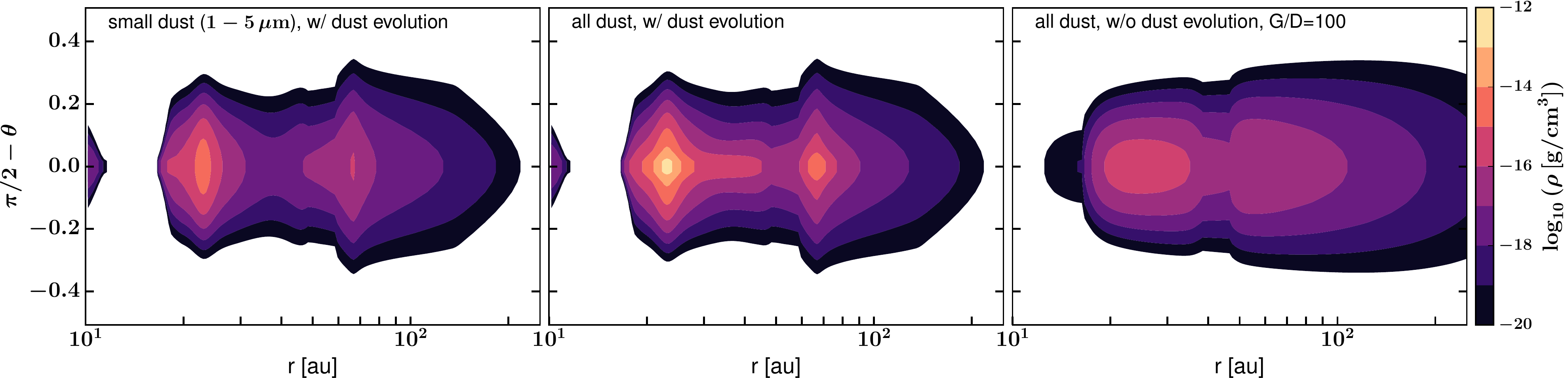}
	}
	\caption{Vertical disk density structure assumed for the radiative transfer calculations following Eq.~\ref{eq:volumedens} for an outer planet mass of 0.7\,M$_{\mathrm{jup}}$. The cumulative density distribution for small dust grains only from $1-5\,\mu$m (\textit{left}), for all dust grains (\textit{middle}) and for the simplified parametric approach (\textit{right}) are shown. Note that the radial scale is logarithmic for a better visualization.}
	\label{fig:volumedens1}
\end{figure*}

\noindent where $R=r\sin(\theta)$ and $z= r\cos(\theta)$ are cylindrical coordinates and $\theta$ the polar angle. The opacity calculation of each grain size bin takes into account porous spheres with a dust mixture composed of astronomical silicates \citep{draine2003}, carbonaceous material \citep{zubko1996}, and water ice \citet{warren2008}. The fractional abundances of 7\%, 21\% and 42\% (amount of vacuum is 30\%) are adopted from \citet{ricci2010}. The temperature structure of each dust grain size is determined with a Monte Carlo radiative transfer simulation and synthetic scattered light images are computed including the full treatment of polarization. Mie theory is used to compute the Mueller matrix elements. These images are convolved by an elliptical Gaussian PSF (0\farcs034 $\times$ 0\farcs041), chosen to mimic the angular resolution of our SPHERE observations, and each pixel is multiplied with the square of its distance to the star to compensate for the stellar illumination drop off with distance. For the synthetic mm observations, we consider a beam size of 0\farcs3 $\times$ 0\farcs2 \citep{fedele2017}.

For comparison reasons, we additionally perform simplified models by neglecting the self-consistent dust evolution, that is, the dust growth, and its dynamics. However, there are mm grains in these models, so significant evolution has taken place here as well. In this second approach, we consider the same initial gas density profile perturbed by the two giant planets and assume a fixed gas-to-dust ratio exploring the range from 50 to 100, a simple approach typically used to compare scattered light images with hydrodynamical simulations of planet-disk interaction \citep[e.g.][]{dong2017}. This approach is valid as long as the micron-sized particles are well coupled to the gas and no self-consistent dust settling is included. However, these simplified models are expected to differ from dust evolution models because several processes, such as growth and fragmentation can change the dust distribution in the disk, which at the same time changes the dynamics, in particular when pressure maxima are present (cf. Fig.~\ref{fig:dustdensity1}). In these simplified models, an average opacity is used considering a power-law distribution for the grain size, where the number density follows $n(a)\,\propto\,a^{-3.5}$ with $a_{\mathrm{min}}=0.01\,\mu$m and $a_{\mathrm{max}}=1\,$mm. For all models, we consider the following stellar parameters $T_{\rm{eff},*}$ = 8400\,K, $M_{*}$ = 1.65\,M$_{\odot}$ and $R_{*}$ = 1.5\,R$_{\odot}$, hence 10\,L$_{\odot}$ \citep{dunkin1997,blondel2006,fedele2017}. The stellar luminosity adopted in \citet{fedele2017} is based on the new distance estimate from Gaia (d=117\,pc). For the stellar spectrum a Kurucz spectrum of a star with metallicity [Fe/H]=0 and a surface gravity of $\mathrm{log}\,g=4.5$ is taken into account \citep[cf.][]{folsom2012}.

In addition, as we do not know the shape of the innermost disk (masked by the coronagraph), and because the presence of a tiny amount of dust could alter the brightness signal close to the inner peak, we set the dust density to a floor value within 10\,au and apply a smooth Gaussian taper to create a rounded inner rim for the inner ring. This step is especially needed in the simplified models, because in the models with dust evolution included, most of the dust particles are filtered out and trapped at the outer edge of the gap opened by the innermost planet, and hence in these models the inner disk is anyway almost empty of grains. However, we note that a tiny inner disk exists as presented in \citet{lazareff2017}.

\begin{figure*}
	\centering
	\centerline{
		\includegraphics[width=\columnwidth]{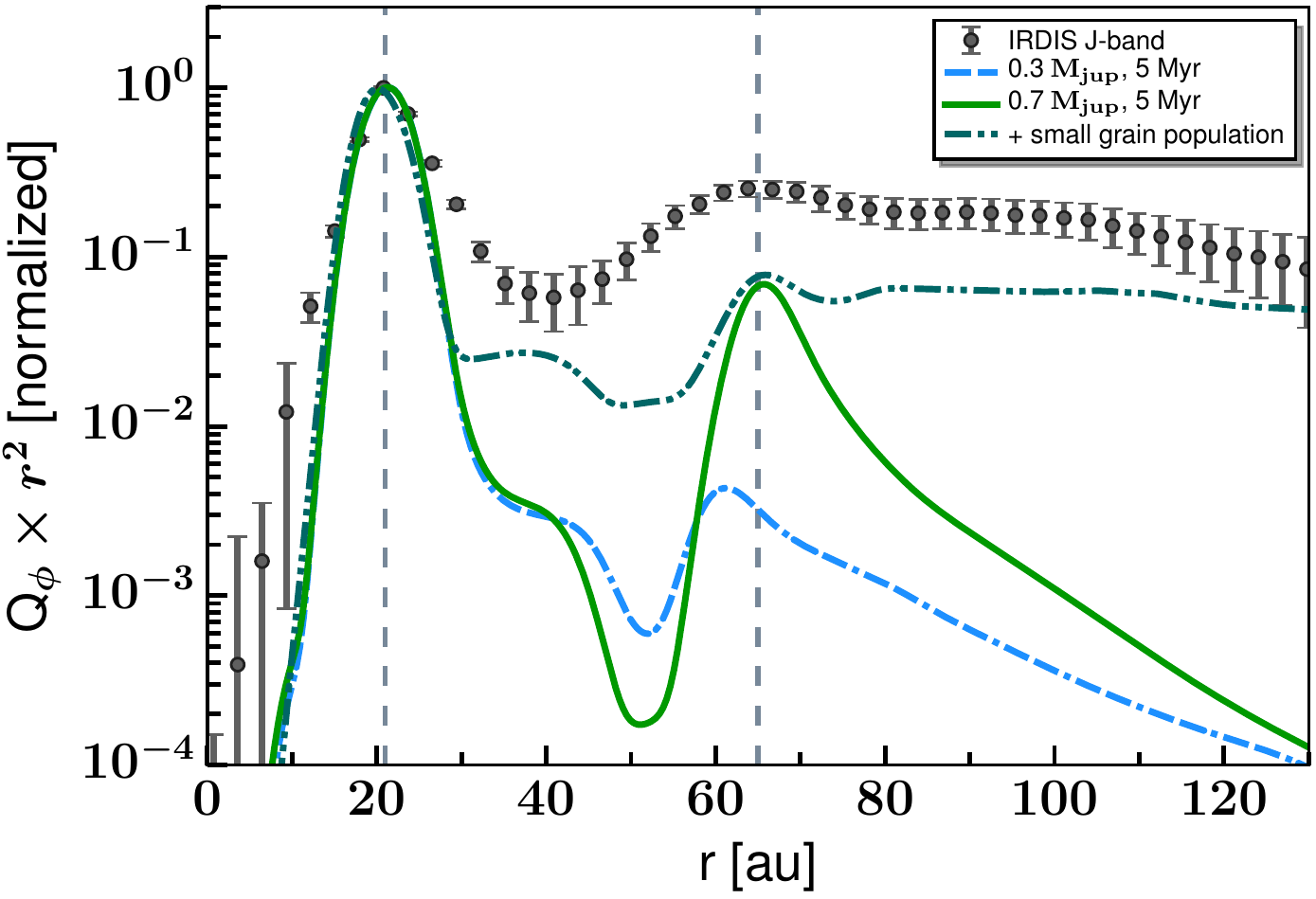}
		\includegraphics[width=\columnwidth]{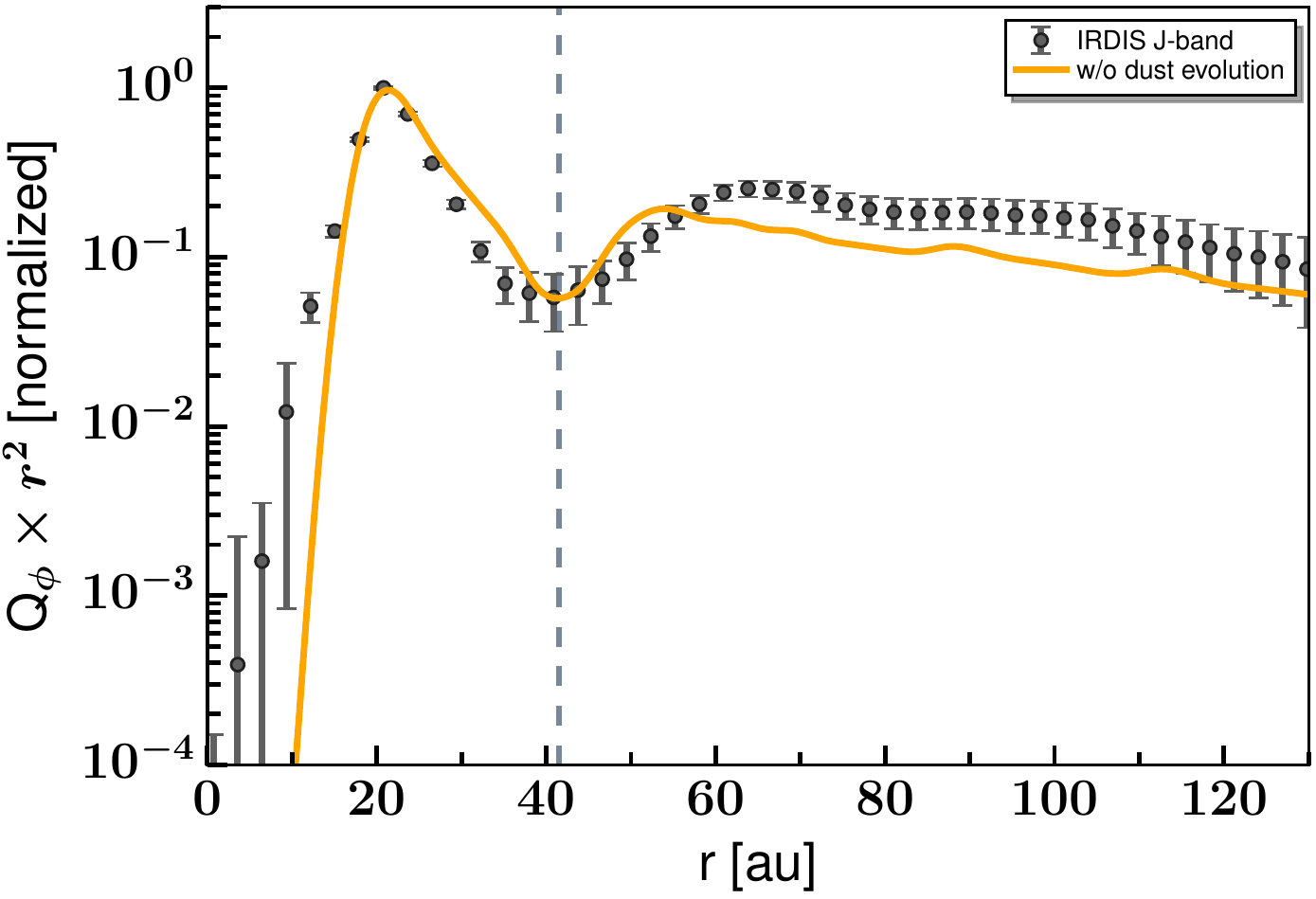}
	}
	\caption{Comparison of surface brightness radial profiles of the $Q_{\phi} \times r^2$ model image at $J$-band. The values are scaled by the square of the distance from the central star in order to compensate for the fall-off of the stellar irradiation. Models with self-consistent dust evolution are shown in the left panel and with a simplified parametric grain size distribution in the right panel. The vertical dashed lines mark the positions of the brightness peaks (\textit{left}) and the location of Gap\,\#2 (\textit{right}), respectively.}
	\label{fig:rtprofiles}
\end{figure*}

\subsection{Results} 
\label{subsec:dert_comp}
Figure~\ref{fig:dustdensity1}, left, shows the dust density distribution after 5\,Myr of evolution for planet masses of 3.5\,M$_{\rm jup}$ and 0.7\,M$_{\rm jup}$ located at 14\,au and 53\,au, respectively. A pressure bump is formed at the outer edge of each gap, which acts as a particle trap and helps to reduce the radial drift. The higher the mass of the planet the more efficient is the trapping and the higher is the mm flux \citep{pinilla2012,pinilla2015a}. This trend is also seen in Fig.~\ref{fig:dustdensity1}, right, where the second planet has a lower mass (0.3\,M$_{\rm jup}$) leading to less efficient trapping there. Although the trapping of mm grains is effective, the values for the planet mass (at the considered disk turbulence) are chosen such that the small grains are not fully filtered out. Note that there is a degeneracy between the choice of disk mass, temperature, $\alpha-$turbulence, and planet mass, thus, we do not claim to infer mass limits for potential planets. The density contrast between the two rings also depends on whether the planets formed simultaneously or sequentially \citep{pinilla2015b}, or whether they migrate. Applying Eq.~\ref{eq:volumedens} to the dust density distributions results in the vertical density structure illustrated in Fig.~\ref{fig:volumedens1}, left and middle panels, and \ref{fig:volumedens2}. While the small grains are distributed radially over the disk extension and all the way up to the disk surface layers according to their dust scale height, the large grains are concentrated at the pressure bump regions close to the midplane. For comparison, Fig.~\ref{fig:volumedens1}, right panel, shows the vertical density structure of our simplified approach, where no grain growth model is involved and a fixed gas-to-dust ratio of 100 is considered. In this case a larger amount of dust is still present within the two gap regions and in the outer disk because the dust radial drift is neglected.

\subsubsection{Scattered light}
\label{subsubsec:comp_sphere}

\begin{figure*}
	\centering
	\centerline{
		\includegraphics[width=\textwidth]{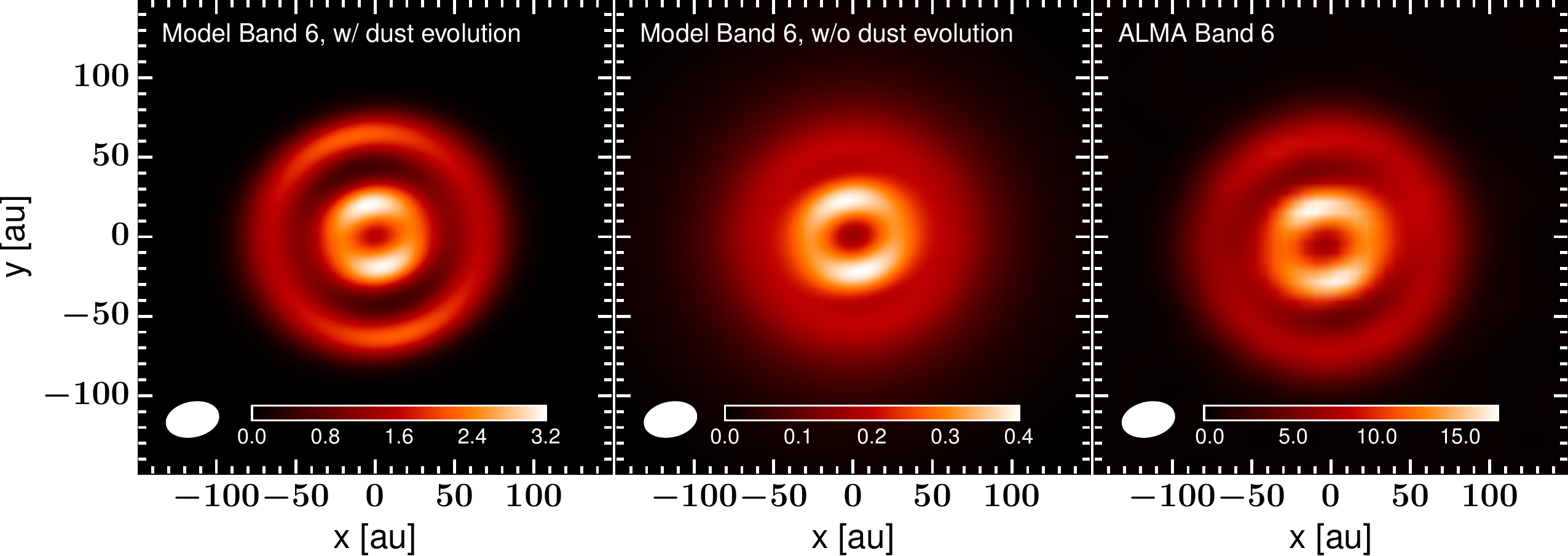}
	}
	\caption{Simulated intensity images of HD\,169142 for ALMA Band 6 (1.3\,mm) based on the radiative transfer models. The images are smoothed to the same angular resolution as in \citet{fedele2017}. The intensity units are mJy/beam.}
	\label{fig:rtband6}
\end{figure*}

Figure~\ref{fig:rtprofiles} shows the radial surface brightness profiles of the $Q_{\phi} \times r^2$ model images, compared to the observational radial profile (also shown in Fig.~\ref{fig:cut}). All profiles are normalized to the peak flux of the inner ring (Ring\,\#2). The brightness of this ring in scattered light is highest because of the geometry of the disk scattering surface. The incidence angle of stellar radiation is steepest here so that the disk receives and scatters most light per unit surface area. This is due to density effects given the large amount of dust there and the temperature profile. The curves in the left panel are based on our dust evolution modeling approach. The solid green curve representing the model with a higher outer planet mass well reproduces the two main ring locations observed with SPHERE, but there is a discrepancy for the brightness contrast between the rings. Although the overall width of Gap\,\#2 matches the observations, a gap much deeper and with sharper edges than observed is produced by our model. Reducing the mass of the outer planet helps to create a slightly shallower outer gap edge (dashed-dotted blue curve). However, in this case the amount of dust that is trapped in the outer disk is lower, and consequently, the brightness of Ring\,\#1 decreases. In addition, the peak of this outer ring slightly moves inwards when the planet mass is reduced, differing from the observations. In both cases, the outer disk in our models appears too faint in scattered light, as all dust grains originally located in the outer disk have already grown and drifted inwards, even after an evolutionary time of 0.5\,Myr (see~Appendix, Fig.~\ref{fig:dustdensity2} and dashed olive curve in Fig.~\ref{fig:rtprofiles_10Myr}).

The right panel (orange line) illustrates a good match for our simplified approach that ignores dust growth and fragmentation processes and assumes a power law for the dust size distribution. The gap location between the two main rings can be reproduced well when moving the second planet position further in from 53\,au to 42\,au. The reason why the planet position needs to be different between the dust evolution models and the simplified approach is the following. The gap in small grains is similar in shape as in the gas surface density, since the small grains are well coupled to the gas. Contrarily, in the dust evolution models there is a dominating peak of emission at the pressure maximum, which is further out from the outer edge of the gas gap and where small grains are continuously reproduced by fragmentation due to turbulent motions \citep{pinilla2012}. The shallowness of the gap better matches compared to the dust evolution approach. However, a simultaneous fit of gap depth and ring positions is not possible either. This trend is also seen in the HD\,169142 model-fitting results by \citet{monnier2017}, where either the gap depth or the outer ring position is off compared to the GPI $J$- and $H$-band profiles.\\

\noindent \textbf{Mixed midplane-surface dust models}\\
At this point of the analysis it seems that the model without dust evolution does a significantly better job in reproducing the SPHERE scattered light observations, which is, however, not the case for the mm dust continuum as demonstrated later in Sect.~\ref{subsubsec:comp_alma}. One has to keep in mind that dust evolution assumptions are developed for the disk midplane, where dust growth is quite efficient due to the high densities. Since the dust evolution models are only 1D, the vertical disk structure chosen influences the situation at larger height. More precisely, the coagulation equation itself is not only calculated in the midplane, but it averages the processes with presumed weights over the vertical structure. Then, it assumes that the size distribution at a given location develops as a whole, followed by a redistribution of the grains. There is also a reservoir of small grains produced, which are going through the growth \& fragmentation cycle. It might be that the vertical exchange in the dust evolution is not working properly and that there are small grains at the disk surface that do not grow quickly at high altitude where the densities are lower. If this population at the top layer is indeed isolated, its coagulation compared to the midplane situation will also be on a different time scale. Moreover, charging effects could play a role for dust evolution processes at the disk surface, which would keep particles very small.

Hence, as a test we introduce a new population of small grains (0.01--0.5\,$\mu$m) that follow the initial gas density distribution of the dust evolution model with a mass fraction of $8 \cdot 10^{-4}\,\Sigma_{\mathrm{disk,gas}}$. The mass in the other size bins is reduced correspondingly to keep the same dust mass as for the original dust evolution simulation. This model is displayed in Fig.~\ref{fig:rtprofiles} with the dashed-dotted dark green line. It helps to decrease the gap depth and to increase the scattered light in the outer disk.

We note that our models, that do not contain any dust inside 10\,au, provide a good match to the SED for wavelengths longer than 10 microns, that trace the outer disk. However, the addition of a small inner belt between $\sim$0.05 and 0.09\,au allows to reproduce the NIR excess as well. To not be seen in our scattered light model predictions, any dusty material in the inner disk must be confined within $\sim$0.09\,au.

\subsubsection{Millimeter dust continuum emission}
\label{subsubsec:comp_alma}
Figure~\ref{fig:rtband6} shows synthetic mm continuum images at 1.3\,mm for our two representative models (solid lines in Fig.~\ref{fig:rtprofiles}) alongside the ALMA data from \cite{fedele2017}. The left panel considers our self-consistent dust growth model. This results in an inner dust cavity, an inner ring between $\sim$15 and 35\,au (0\farcs13 and 0\farcs3) and an outer ring between $\sim$55 and 80\,au (0\farcs47 and 0\farcs68), with a gap in between. Both the inner cavity and the gap are depleted in mm-sized dust particles. Our model is qualitatively consistent with the ALMA dust continuum image showing rings at $\sim$20--35\,au (0\farcs17--0\farcs28), and $\sim$56--83\,au (0\farcs48--0\farcs64). This agreement supports the view that the efficient dust trapping scenario by means of the two giant planets may be at work in HD\,169142. The relatively sharp outer edge of the continuum map gives further evidence of large dust grains radially drifting inwards \citep[cf.][]{birnstiel2014,facchini2017}. For completeness the synthetic image for our simplified fixed gas-to-dust ratio model without dust evolution treatment is also shown. The clear depletion of dust particles within the gap region is not seen in this case. Furthermore, the outer ring is more extended and both the inner and outer edge are less well defined leading to a fuzzier overall disk structure. We note that the flux is under-predicted in both model scenarios compared to the actual ALMA measurement.

\section{Discussion}
\label{sec:disc}
 
\subsection{Fragmentation and trapping efficiency} 
\label{subsec:disc_frag}
The main reason for the discrepancy between the dust evolution models and the simplified approach is the fact that in the dust evolution models, micron-sized particles efficiently grow to larger sizes already at early times of evolution. The growth changes their coupling to the gas (i.e. their Stokes number) and hence their dynamics (e.g., their drift velocities increase when they grow). In these models, small grains are continuously reproduced thanks to destructive collisions that can occur because of turbulent motions and radial drift. In the particular case of two planets embedded in the disk presented in this study, the radial drift is reduced at the pressure bumps and fragmentation occurs due to turbulence that replenishes these regions with small grains. These small grains are more affected by turbulent motions, they are more difficult to trap, and thus they can be dragged along with the gas. For this reason, a little amount of micron-sized particles can still flow through the gap (cf. Fig.~\ref{fig:dustdensity1}). This amount of small grains is, however, not enough to reproduce the observed surface brightness profile inside the gap (as it is in the case of a constant gas-to-dust ratio). For sub-micron grains this amount would be significantly higher, and these smaller grains scatter more efficiently in our direction, too. A possible solution for this discrepancy is to make fragmentation more efficient, for example by increasing the turbulent motions of the grains, that is increasing the $\alpha$-viscosity. However, when $\alpha$ increases, a more massive planet is needed to open a gap \citep{crida2006}, which can lead to a new discrepancy with the gap width. Moreover, the higher the turbulent motions, the more difficult it is to trap the mm-sized particles, because of the high dust diffusion that allows particles to escape from pressure bumps \citep[][]{dejuanovelar2016}. An alternative to have less growth and more fragmentation, is to decrease the maximum fragmentation velocity threshold of particles, which mainly depends on the grain composition and its structure. Nonetheless, while having more fragmentation might help to increase the surface brightness inside the gap, this can also lead to less dust trapping, which can create differences with the current mm-observations. Note that decreasing the initial minimum grain size in our current dust evolution models would not help to have a better match with observations because small grains quickly grow regardless of their initial size.

\subsection{Dust evolution as a function of $z$}
\label{subsec:disc_mix}
Following the analysis in Sect.~\ref{subsubsec:comp_sphere} it becomes clear that the assumptions in current 1D dust evolution models are tuned for coagulation processes happening in the disk midplane. The evolution of gas and dust is modeled in a vertically integrated way assuming a steady-state disk model, although there might actually be a strong dependency on the vertical disk height. Hence, our treatment of vertical exchange and vertical settling might also be not accurate. For instance, if there is indeed a population of small grains isolated at the top layer, this would suggest very weak turbulence. This provokes quite efficient settling, even for small grains. What might work is a population of small, charged grains that is kept from settling for example by magnetic fields. Hence, these small grains would be unaffected by efficient dust growth and could be permanently present at the disk surface. As a consequence, scattered light detections at optical and NIR wavelengths would be not effected by significant dust growth. 
Contrarily, the surface layers might have higher turbulence, which is expected because they are hotter and highly ionized \citep[e.g.,][]{dzyurkevich2013}. However, with only higher turbulence, grains are also mixed downwards and get into contact with the lower turbulence regions deeper in the disk, where they would settle and take part in the coagulation down there. Thus, a locally higher turbulence is not a way to isolate grains, it is a way to move them faster. Consequently, it could be that the gas velocities at very high altitude are such that fragmentation also works in a thin surface layer to locally replenish the reservoir of even sub-micron grains.

\subsection{Mass of gap-opening planets}
\label{subsec:disc_mass}
As shown in Sect.~\ref{subsec:dert_comp}, planets with masses of 3.5\,M$_{\rm jup}$ and 0.7\,M$_{\rm jup}$ located at 14\,au and 53\,au are needed in our dust evolution models in order to create effective pressure bumps that trap particles at the location of the rings seen in scattered light. Note that these values are compatible with the mass detection limits derived from contrast curves in total intensity SPHERE IRDIS and IFS data. The minimum planet mass limit in our model is chosen such that the planet perturbs the gas profile and efficient trapping can be generated (M$_{\mathrm{p}} \gtrsim 0.3\,M_{\rm jup}$). While a planet mass of 0.3\,M$_{\rm jup}$ is too low to clear the full extent of Gap\,\#2, the 0.7\,M$_{\rm jup}$ planet is able to reproduce the gap width. The presence of multiple planets below this mass, whose gaps overlap, is an alternative possibility \citep[e.g.,][]{dodson2011}. Numerical studies have shown that less massive planets do not open a gap in the gas, but effectively open a gap in the dust \citep{paardekooper2004,paardekooper2006,picogna2015,rosotti2016,dipierro2016,dipierro2017}. The gas azimuthal velocities can be perturbed such that the drift velocities of the particles are reduced, leading to a traffic jam effect without creating local pressure maxima. In addition to the gas viscous forces, \citet{dipierro2016} and \citet{dipierro2017} also include the contribution from the tides of an embedded planet and show that a low-mass planet can open a gap in the dust only, if the tidal torque exceeds the drag torque outside the planetary orbit. In this scenario, a shallow gap can be carved out, but it is rather unlikely that this effect can create the strong rings in the distribution of small and large grains in HD\,169142. It should be tested whether a combination of pressure bumps, self-consistent dust evolution and the consideration of disk-planet tidal interactions can lead to a coherent picture for the gap and ring appearances.

Pioneering studies from \citet{kanagawa2015}, \citet{rosotti2016} and \citet{dong2017} look at the inverse problem, meaning to derive planet masses from observed gap profiles. For this method a number of assumptions about the disk structure and dynamics are made when simulating the gap shape and deriving the correlation with planet mass. The mass of the putative second planet in our HD\,169142 model is consistent with the numerical analysis by \citet{dong2017}, who derived disk and planet properties based on the morphology of gaps in NIR scattered light images. They estimate a mass between 0.2--2.1\,M$_{\rm jup}$ for an $\alpha$-viscosity varying from 10$^{-4}$--10$^{-2}$. \citet{kanagawa2015} suggest a mass $\gtrsim$\,0.4\,M$_{\rm jup}$ by measuring the gap depth in VLA 7\,mm data. Although this is principally consistent with the other estimates, a measurement based on mm data only is complicated due to dust/gas coupling effects \citep{rosotti2016}. This makes an exact definition of the gap width difficult and its value depends on the disk lifetime. As discussed in \citet{rosotti2016} a more robust indicator of the planet mass from (sub-)mm images is the location of the bright ring tracing the gas pressure maximum. This is the reason why we intend to reproduce the ring positions rather than the gap locations with the modeling approach in this paper. The inclusion of dust growth and fragmentation processes would certainly change the conclusions from \citet{rosotti2016} and \citet{dong2017} as dust evolution dynamics affects the gap depth, the slope of the gap edges, the position of the rings and their contrast.
 
\subsection{Dust evolution timescale}
\label{subsec:disc_time}
All our results based on the dust evolution modeling consider a dust evolutionary timescale of 5\,Myr, and that the giant planets embedded were formed simultaneously. The disk and planet age can affect the appearance of the radial profiles in polarized intensity at the NIR (see Fig.~\ref{fig:rtprofiles_10Myr} in the Appendix) and in total intensity at mm wavelengths. On the one hand, the outer ring (Ring\,\#1) becomes narrower at longer times of evolution, which produces a rather sharp outer disk edge and shifts it towards smaller radii. This in turns lowers the brightness signal in the outer disk. While this is consistent with the mm data, the amount of small dust particles decreases with time, and the NIR observations cannot be reproduced. If longer times of evolution are taken ($\sim$10\,Myr, which is consistent with the revised age of the system), there would be a higher discrepancy between the dust evolution models and the NIR observations unless additional trapping mechanisms play a role all across the disk. In contrast, at very early timescales of 0.1--0.5\,Myr, when all grains are not yet at the pressure maxima, the wide gap (Gap\,\#2) remains shallower. On the other hand, analogous to the sequential planet formation scenario presented in \citet{pinilla2015a}, it could be also possible that the outer planet forms earlier than the inner planet (or vice versa). This can affect the amount of dust in both traps and adjust the contrast between the two rings. However, we do not have any constraint on whether the two planets have been forming at the same time or consecutively. Together with the uncertainty when the putative planets have been forming at all, this means that the dust evolution after 5\,Myr could be still a good proxy for the situation in the HD\,169142 system.

\subsection{Gaps and rings in the context of ice lines}
\label{subsec:disc_ice}
\begin{figure}
	\centering
	\centerline{
		\includegraphics[width=\columnwidth]{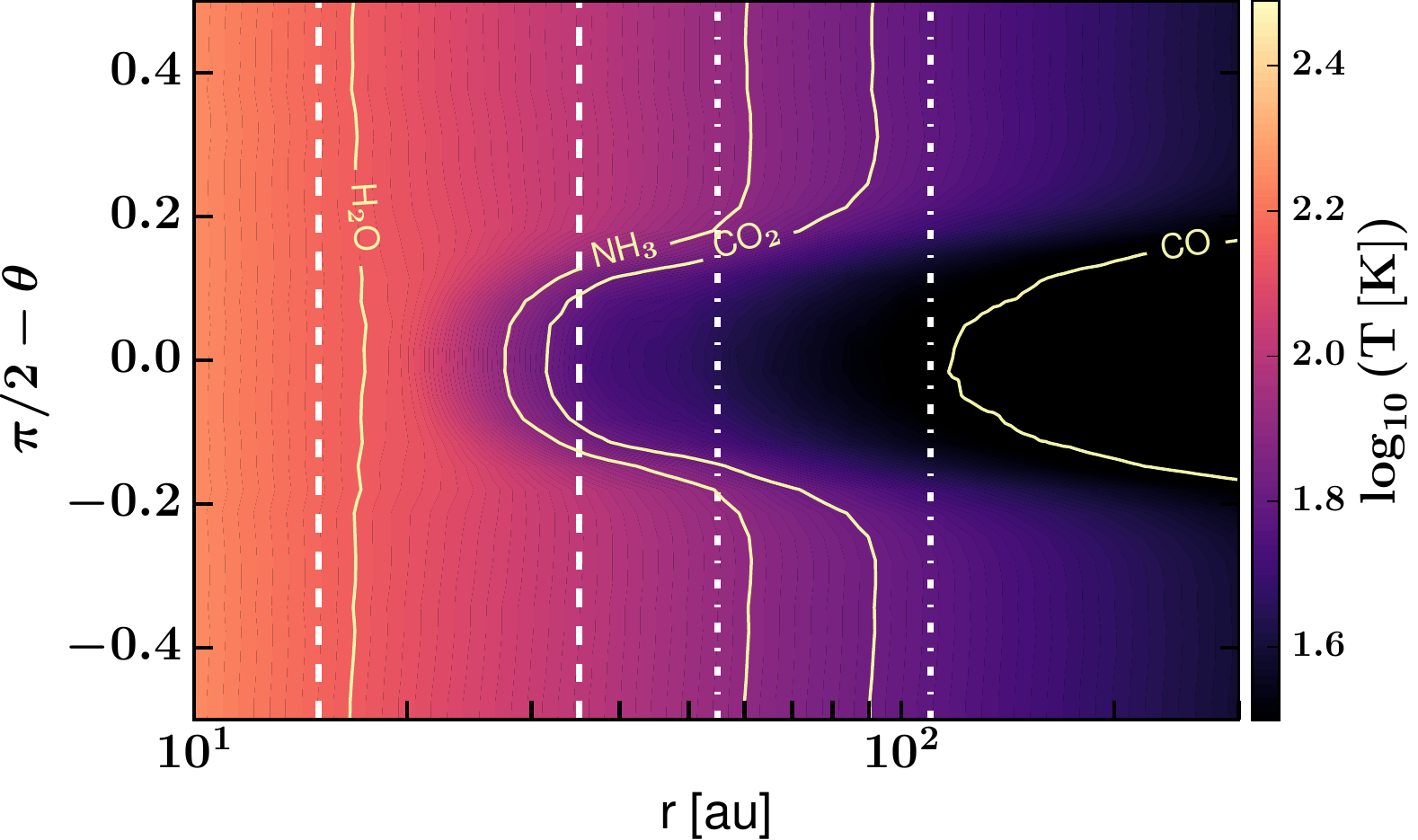}
	}
	\caption{Temperature map as a function of radius and polar angle in spherical coordinates for the simplified model. The ice lines for H$_2$O, NH$_3$, CO$_2$ and CO are indicated with yellow contour lines. The vertical dashed lines illustrate the edges of the two bright rings in mm dust emission.}
	\label{fig:icelines}
\end{figure}

Although observational signposts of embedded planets are the most widely used explanation to interpret ring structures in disks, the relation to ice lines of various materials is another possible scenario \citep{zhang2015,okuzumi2016}. Ice lines of different volatile species can significantly affect the dynamics of dust evolution processes including growth and fragmentation, which in turn has an effect on the observational appearance of rings and gaps at different wavelengths \citep{pinilla2017}. The freeze-out temperatures of main volatiles, such as water (H$_2$O), ammonia (NH$_3$), carbon dioxide (CO$_2$) and carbon monoxide (CO), are estimated to have average values of $\sim$142\,K, $\sim$80\,K, $\sim$66\,K and $\sim$26\,K, respectively \citep{zhang2015}.

Our radiative transfer models show that the midplane temperatures at the inner ring position are such that the H$_2$O and NH$_3$/CO$_2$ ice lines are located close to the inner and outer edges of this ring at mm emission, respectively. It is recognizable that the gap at scattered light lies between the ice lines of H$_2$O and CO$_2$ when comparing the surface layer temperatures with the volatile freeze-out temperatures. Thus, the H$_2$O, NH$_3$ and CO$_2$ ice lines nearly coincide with the scattered light ring positions. Furthermore, the CO ice line at the midplane is located at $\sim$110\,au, which is close to the outermost gap at $\sim$85\,au, consistent with DCO$^{+}$(3--2) and C$^{18}$O(2--1) ALMA observations presented in \citet{macias2017}. In Fig.~\ref{fig:icelines}, there is an uncertainty for the specific location of these ice lines, which depends on the freezing temperatures that we assume, and for the CO ice line, it could be between $\sim$95 and 145\,au. The current observations suggest that the accumulation of large dust grains close to the CO ice line is a possible mechanism to explain the origin of this outermost gap.

\subsection{Shadowing effects and time variability}
\label{subsec:disc_shadow}
We note that because of the modeling procedure (analytical gas profile coupled with 1D dust evolution) the observational signatures presented in this paper are always azimuthally symmetric. This is for example not necessarily true for massive enough planets for which an eccentric gap and vortex formation at its edge are expected \citep[e.g.,][]{ataiee2013}. As mentioned in Sect.~\ref{sec:obs} there are significant asymmetries both along the inner and outer ring regions in polarized intensity. It is noticeable that the maximum polarization of the outer ring is along the minor axis. This is opposite to several other disks showing a brighter polarized intensity along the major axis, expected due to the polarization efficiency being highest for 90\dg\, scattering in the Rayleigh and Mie scattering regime. A significant scattering angle effect is also likely not to be expected at the low inclination of HD\,169142. \citet{momose2015} invokes corrugations of the scattering surface in the outer region as a possible origin. Alternatively, asymmetries in the outer disk emission can be caused by shadowing of the inner disk region. For HD\,169142 there is a slightly inclined inner disk at sub-au distance \citep[i=21\dg, PA=100--130\dg,][]{lazareff2017}, which is known to be variable and might contain an extended dust envelope as suggested by \citet{wagner2015}. Azimuthal brightness variations in the inner ring of the scattered light could be caused for example by perturbations by a protoplanet, by optical depth variations through the suggested dust envelope, by accretion flows or turbulence in the inner disk. The local brightness enhancements along the innermost scattered light ring at different time epochs are discussed in \citet{ligi2017}. The azimuthal inhomogeneities in the inner ring (Ring\,\#2) could in turn cause radial shadowing on Ring\,\#1 and the remaining outer disk. The pace of variations in the illumination pattern of the outer disk depends on the precession timescale of the inner disk material. Given that there is no apparent difference in the rings' brightness asymmetries in the three observational data sets in polarized intensity (Subaru/HiCIAO: \citealt{momose2015}, Gemini South/GPI: \citealt{monnier2017}, VLT/SPHERE: this paper) that span a time period of three years, the shadowing scenario for the outer disk seems rather unlikely. However, it cannot be ruled out either, as the precession time scale for the inner disk, for example in the context of a hypothetic star-companion system, can be rather long (several hundred to thousand years).

\section{Conclusions}
\label{sec:concl}

In this paper, we present scattered light observations of the protoplanetary disk around the Herbig Ae star HD\,169142 obtained with VLT/SPHERE at $J$-band, and compare our results with recent ALMA data of this target. Together with TW\,Hya, HD\,163296 and HD\,97048, it is one of a handful of disks around young stars that have been observed at very high-angular resolution at NIR and mm wavelengths, where in each case both images show  similar sub-structures even if their scales differ, but also show different morphologies. For HD\,169142 we confirm the previous detection of two ring-like features separated by a wide gap, and an additional inner gap, and report on the marginal detection of a third gap in the outer disk as well as azimuthal brightness variations along both rings. We present azimuthally symmetric radiative transfer models based on planet-disk interaction processes that account for the main observational features, and discuss the influence of dust evolution and particle trapping on the gap and ring properties. We place our findings in the context of planet masses inferred from the gap-opening process and ice line chemistry. Our measurements and modeling results suggest the following:

\begin{enumerate}
	\item The location and width of the gap, as well as the peak positions in polarized scattered light of HD\,169142 are reproduced with our model based on dust evolution processes when two giant planets of 3.5 and 0.7\,M$_{\rm jup}$ are embedded in the disk. The observed gap, however, possesses a shallower outer flank than expected for planet-disk interaction signatures. There is also a significant discrepancy for the gap depth as micron-sized particles rapidly grow in the presence of pressure bumps. Small grains distributed all over the disk, wherever there is gas, as in our simplified approach decreases the gap depth such that there is a good agreement with the observed shape. This also helps to increase the scattered light flux in the outer disk. A more efficient fragmentation by increasing the turbulent motion of dust particles or to adjust the fragmentation velocities could help to overcome this deficit in small grains. Including the contribution from the tides of an embedded planet can lead to a shallower dust gap, in case the planet hypothesis is correct at all.
	\item The assumptions in current dust evolution models are tuned for the disk midplane and the vertical exchange does not work properly, thus, the coagulation timescale might be different at higher disk altitudes. A population of small (sub-)micron-sized grains might exist in the upper surface layers that is unaffected of quick growth due to lower densities and different turbulence there. Thus, the vertical disk structure and its consequence on dust evolution processes also have a significant role for interpreting scattered light images.
	\item In order to obtain a consistent picture with the mm observations, the accumulation of large grains in the dust trap of a pressure bump is needed. This generates the bright emission rings and the sharp outer disk edge as detected in the mm continuum image of HD\,169142. A simplified parameterized dust size distribution is not able to reproduce the high dust depletion factor required.
    \item A scenario with a grain size dependent gap opening that still allows a perturbation in the radial pressure gradient is required. We emphasize that inferring the mass of gap-opening planets from simplified models is degenerate and depends on the choice of the disk mass, temperature, and $\alpha-$turbulence. Constraining planet masses becomes even more uncertain when including more physical processes that are expected to occur in protoplanetary disks, such as grain growth, fragmentation and vertical disk instabilities. 
    \item Observing the total amount of gas and using different techniques that allow us to get better constraints on the grains sizes in disks, such as mm-wave dust polarization \citep{kataoka2015,kataoka2016,pohl2016,yang2016}, may allow to further explain the origin of the gaps and to derive properties of potential embedded planets.
\end{enumerate}

Eventually, 2D dust evolution models are needed in order to have a self-consistent treatment of radial transport and vertical settling, and to consider turbulence changes across the vertical direction of the disk. In principle, we can start to use multi-wavelength analyses, such as the one presented in this paper, to provide feedback on the model assumptions, and to calibrate our understanding of microphysical dust processes (sticking, fragmentation, compact vs. fluffy grains, etc.).

\appendix

\section{$U_{\phi}$ image and polar mapping}

\begin{figure}[!h]
	\centering
	\includegraphics[width=0.45\textwidth]{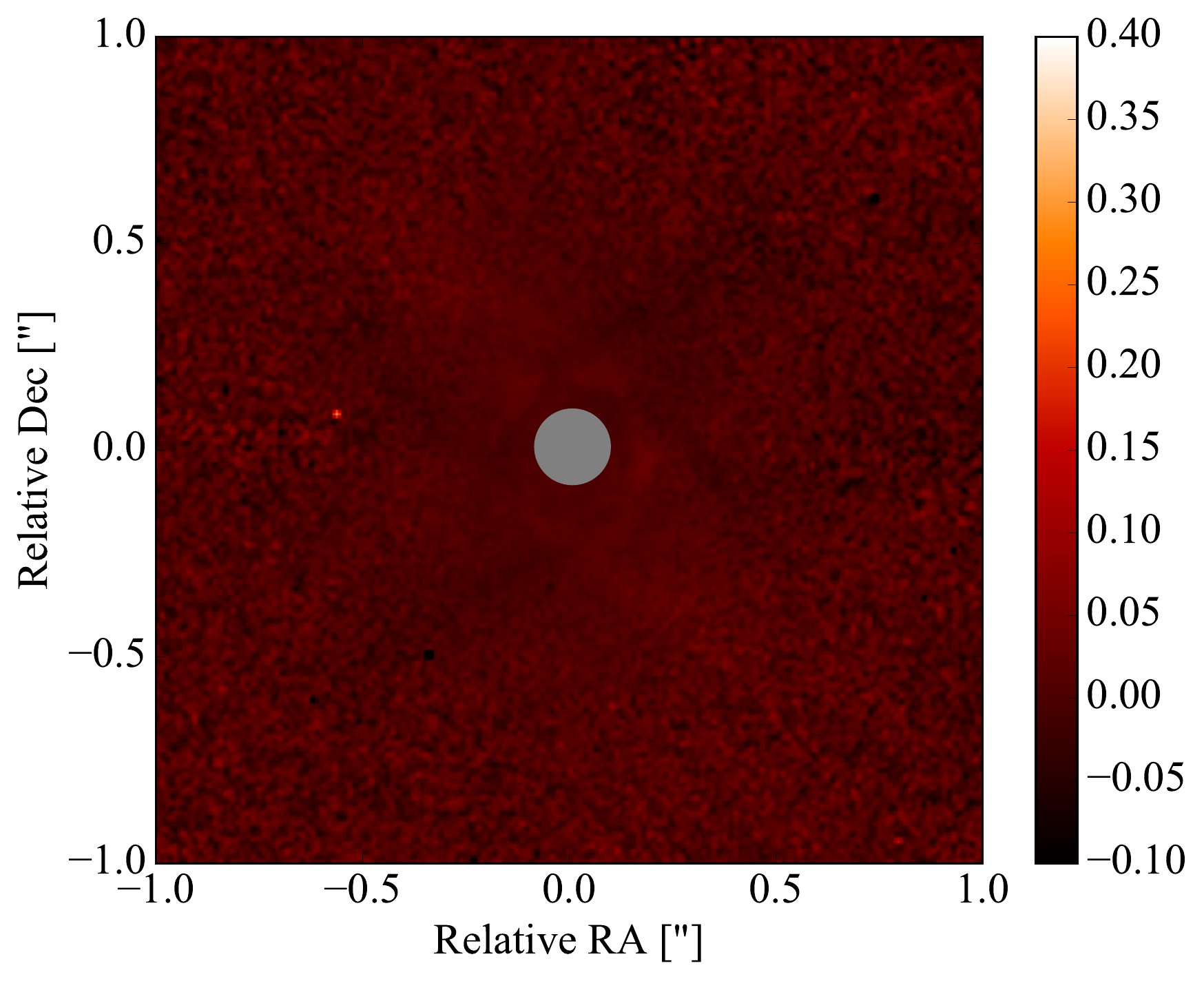}  
	\caption{$J$-band $U_{\phi} \times r^2$ image in linear scale. Each pixel is multiplied with the square root of its distance to the star, $r^{2}$, to compensate for the stellar illumination drop-off with radius. The normalization is simimilar to the $Q_{\phi}$ image, but the dynamical range of the color bar is adjusted. The region masked by the coronagraph is indicated by the grey circle. North is up, East is pointing towards left.} 
	\label{app:uphi}
\end{figure}

\begin{figure}[!h]
	\centering
	\includegraphics[width=0.45\textwidth]{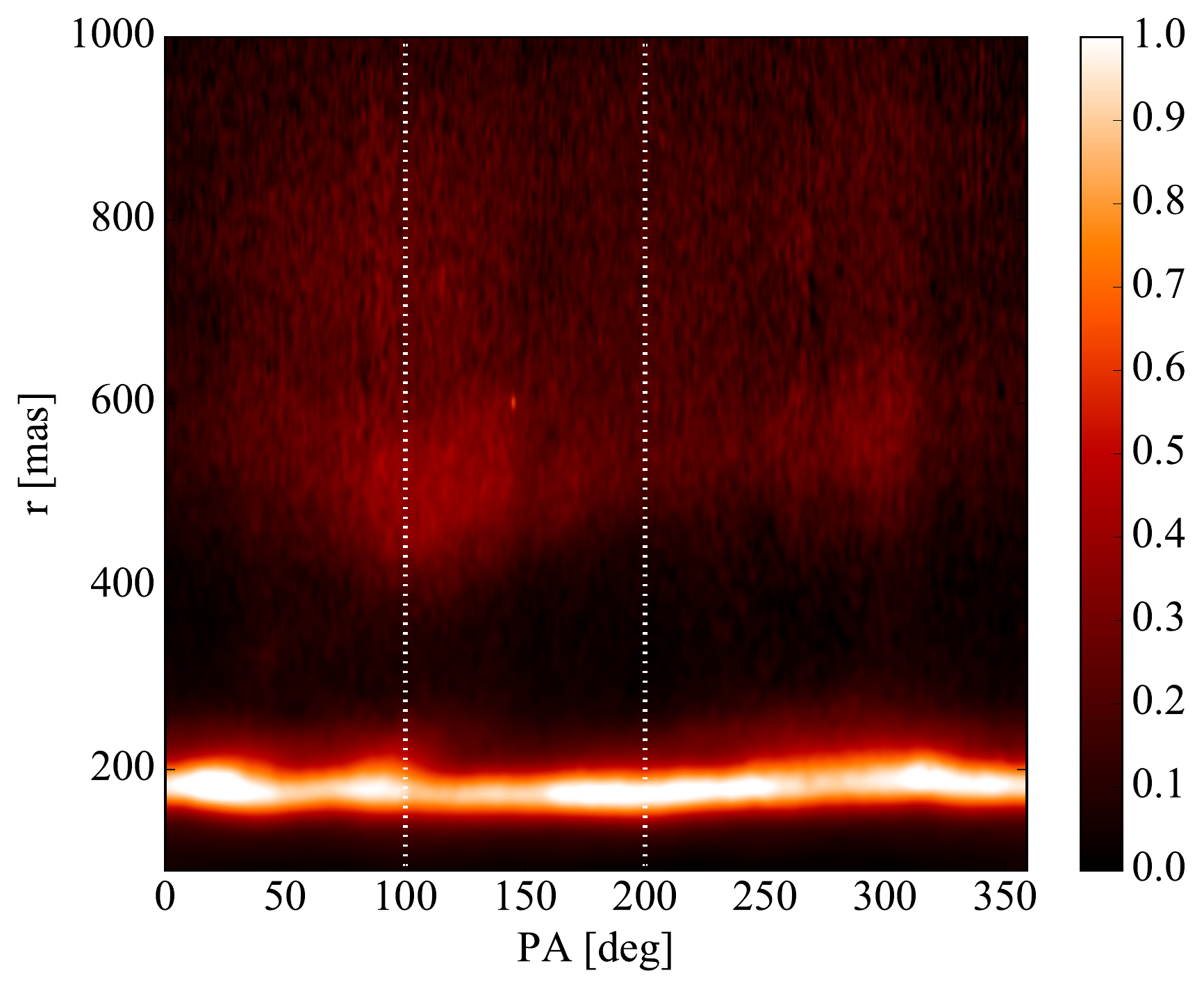}  
	\caption{Polar map of the $Q_\phi \times r^2$ image in linear scale. The vertical dotted lines indicate PA=100\dg\, and 200\dg.} 
	\label{app:polar}
\end{figure}

\newpage
\section{Ellipse fitting} 

\begin{table}[!h]
	\label{tab:ellipsesFixed}
	\begin{center}
	\caption{Ellipse parameters fitted to the two rings and Gap\,\#2 in our scattered light images.}
	 \begin{tabular}{lccc}
	 	 \hline
	 	 \hline
	 	 & Ring\,\#1 & Gap\,\#2 & Ring\,\#2 	\\
	 	 \hline
 		 $\Delta$ RA [mas]		& 28.4$\pm$5.5 & 12.4$\pm$5.3 & 4.4$\pm$2.9 \\
 		 $\Delta$ Dec [mas]		& 18.9$\pm$5.7 & 33.5$\pm$5.4 	& 5.3$\pm$2.8 \\
 		 semi-major axis [mas]	& 536.4$\pm$18.2 & 375.2$\pm$14.5 & 173.8$\pm$9.1 \\
 		 semi-major axis [au]		& 62.8$\pm$2.1& 43.9$\pm$1.7 & 20.3$\pm$1.1 \\
 		 semi-minor axis [mas]	& 522.7$\pm$1.2 & 365.6$\pm$1.5 & 169.3$\pm$2.0 \\
 		 semi-minor axis [au] 	& 61.2$\pm$0.1 & 42.8$\pm$0.2 & 19.8$\pm$0.2 \\
 		 \hline
 		 \end{tabular}
 	\end{center}
	\tablecomments{The position angle of the disk is fixed to 5\,deg and the inclination to 13\,deg. We give the offset of the ellipses from the star position as well as the size of the semi-major and -minor axes for each fitted feature. We also state the size of the semi-major axis in au, since it directly corresponds to the physical radius of the deprojected rings.}
\end{table}

\section{Modeled dust density distribution}

\begin{figure*}[!h]
	\centering
	\centerline{
		\includegraphics[width=\textwidth]{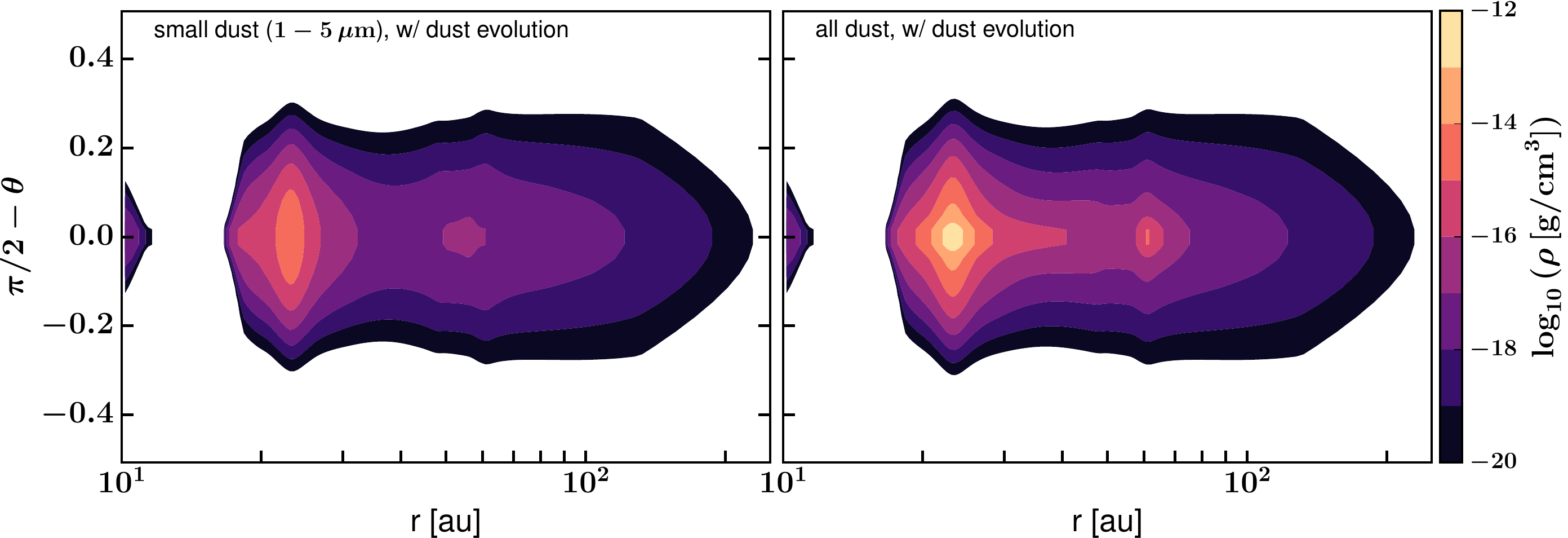}
	}
	\caption{Vertical disk density structure assumed for the radiative transfer calculations following Eq.~\ref{eq:volumedens} for an outer planet mass of 0.3\,M$_{\mathrm{jup}}$. The cumulative density distribution for small dust grains only from $1-5\,\mu$m (\textit{left}) and for all dust grains (\textit{right}) are shown. Note that the radial scale is logarithmic for a better visualization.}
	\label{fig:volumedens2}
\end{figure*}

\newpage
\section{Effect of dust evolution timescale}

\begin{figure*}[!h]
	\centering
	\centerline{
		\includegraphics[width=\textwidth]{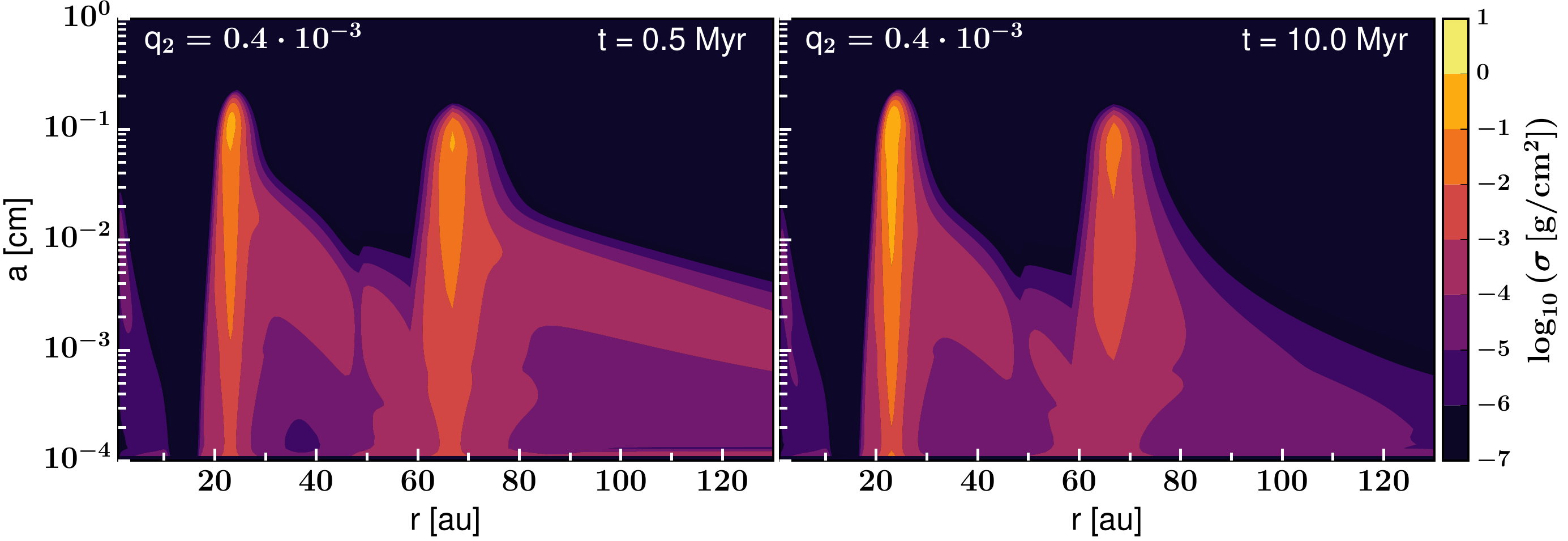}
	}
	\caption{Vertically integrated dust density distribution after 0.5\,Myr (left) and 10\,Myr (right) of evolution, when two massive planets (3.5 and 0.3\,M$_{\rm jup}$) are embedded in the disk at 14\,au and 53\,au, respectively.}
	\label{fig:dustdensity2}
\end{figure*}	

\begin{figure*}[!h]
	\centering
	\centerline{
		\includegraphics[width=0.5\textwidth]{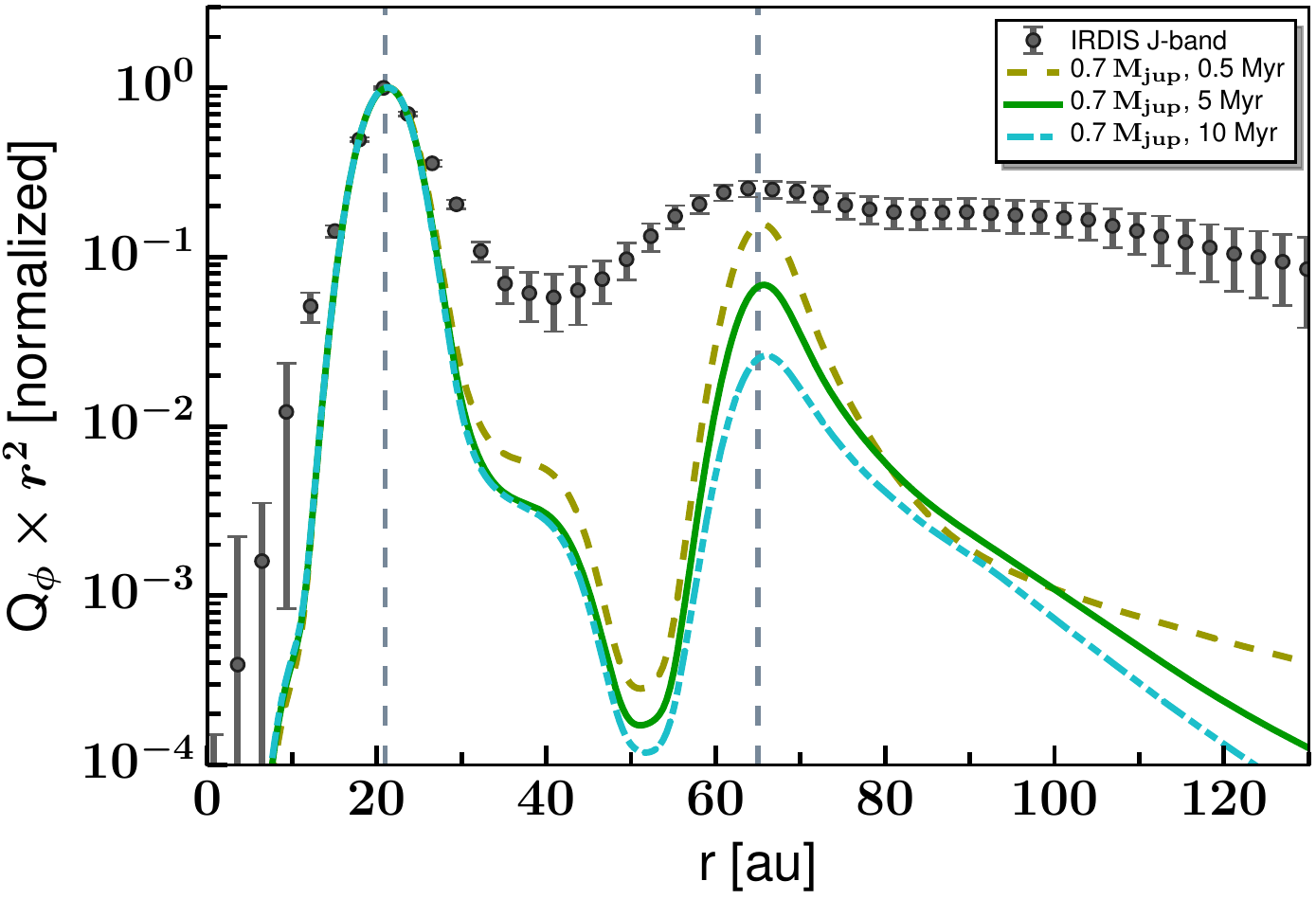}
	}
	\caption{Comparison of surface brightness radial profiles of the $Q_{\phi} \times r^2$ model image at $J$-band for different dust evolution timescales. The values are scaled by the square of the distance from the central star in order to compensate for the fall-off of the stellar irradiation. The vertical dashed lines mark the positions of the brightness peaks.}
	\label{fig:rtprofiles_10Myr}
\end{figure*}

\begin{acknowledgements}
We acknowledge the ESO Paranal Staff for their support during the observations. We thank J.~Monnier for sharing his data. We are thankful to L.~P\'{e}rez, A.~Garufi and G.~Bertrang for the useful discussions about this project. A.P. is supported by the Reimar-L\"{u}st Fellowship of the Max Planck Society and a member of the International Max Planck Research School for Astronomy and Cosmic Physics at Heidelberg University, IMPRS-HD, Germany. We acknowledge financial support from the Programme National de Plan\'{e}tologie (PNP) and the Programme National de Physique Stellaire (PNPS) of CNRS-INSU. This work has also been supported by a grant from the French Labex OSUG@2020 (Investissements d'avenir - ANR10 LABX56). M.B., F.M. and G.P. acknowledges funding from ANR of France under contract number ANR-16-CE31-0013 (Planet Forming Disks). P.P. acknowledges support by NASA through Hubble Fellowship grant HST-HF2-51380.001-A awarded by the Space Telescope Science Institute, which is operated by the Association of Universities for Research in Astronomy, Inc., for NASA, under contract NAS 5-26555. H.A., M.M. and S.P.Q. acknowledges support from the Millennium Science Initiative (Chilean Ministry of Economy) through grant RC130007 and further financial support by FONDECYT, grant 3150643, as well as financial support within the framework of the National Centre for Competence in Research PlanetS supported by the Swiss National Science Foundation. T.B. acknowledges funding from the European Research Council (ERC) under the European Union’s Horizon 2020 research and innovation programme under grant agreement No 714769. D.F. acknowledges support from the Italian Ministry of Education, Universities and Research project SIR (RBSI14ZRHR). Q.K. acknowledges funding from STFC via the Institute of Astronomy, Cambridge Consolidated Grant. SPHERE is an instrument designed and built by a consortium consisting of IPAG (Grenoble, France), MPIA (Heidelberg, Germany), LAM (Marseille, France), LESIA (Paris, France), Laboratoire Lagrange (Nice, France), INAF - Osservatorio di Padova (Italy), Observatoire astronomique de l'Universit\'{e} de Gen\`{e}ve (Switzerland), ETH Zurich (Switzerland), NOVA (Netherlands), ONERA (France) and ASTRON (Netherlands) in collaboration with ESO. SPHERE was funded by ESO, with additional contributions from CNRS (France), MPIA (Germany), INAF (Italy), FINES (Switzerland) and NOVA (Netherlands). SPHERE also received funding from the European Commission Sixth and Seventh Framework Programmes as part of the Optical Infrared Coordination Network for Astronomy (OPTICON) under grant number RII3-Ct-2004-001566 for FP6 (2004-2008), grant number 226604 for FP7 (2009-2012) and grant number 312430 for FP7 (2013-2016). This work has made use of data from the European Space Agency (ESA) mission {\it Gaia} (\url{http://www.cosmos.esa.int/gaia}), processed by the {\it Gaia} Data Processing and Analysis Consortium (DPAC, \url{http://www.cosmos.esa.int/web/gaia/dpac/consortium}). Funding for the DPAC has been provided by national institutions, in particular the institutions participating in the {\it Gaia} Multilateral Agreement.
\end{acknowledgements}

\bibliographystyle{apj}
\bibliography{hd169}

\begin{thebibliography}{}
\expandafter\ifx\csname natexlab\endcsname\relax\def\natexlab#1{#1}\fi

\bibitem[{{ALMA Partnership} {et~al.}(2015){ALMA Partnership}, {Brogan},
  {P{\'e}rez}, {Hunter}, {Dent}, {Hales}, {Hills}, {Corder}, {Fomalont},
  {Vlahakis}, {Asaki}, {Barkats}, {Hirota}, {Hodge}, {Impellizzeri}, {Kneissl},
  {Liuzzo}, {Lucas}, {Marcelino}, {Matsushita}, {Nakanishi}, {Phillips},
  {Richards}, {Toledo}, {Aladro}, {Broguiere}, {Cortes}, {Cortes}, {Espada},
  {Galarza}, {Garcia-Appadoo}, {Guzman-Ramirez}, {Humphreys}, {Jung}, {Kameno},
  {Laing}, {Leon}, {Marconi}, {Mignano}, {Nikolic}, {Nyman}, {Radiszcz},
  {Remijan}, {Rod{\'o}n}, {Sawada}, {Takahashi}, {Tilanus}, {Vila Vilaro},
  {Watson}, {Wiklind}, {Akiyama}, {Chapillon}, {de Gregorio-Monsalvo}, {Di
  Francesco}, {Gueth}, {Kawamura}, {Lee}, {Nguyen Luong}, {Mangum}, {Pietu},
  {Sanhueza}, {Saigo}, {Takakuwa}, {Ubach}, {van Kempen}, {Wootten},
  {Castro-Carrizo}, {Francke}, {Gallardo}, {Garcia}, {Gonzalez}, {Hill},
  {Kaminski}, {Kurono}, {Liu}, {Lopez}, {Morales}, {Plarre}, {Schieven},
  {Testi}, {Videla}, {Villard}, {Andreani}, {Hibbard}, \&
  {Tatematsu}}]{alma2015}
{ALMA Partnership}, {Brogan}, C.~L., {P{\'e}rez}, L.~M., {et~al.} 2015, \apjl,
  808, L3

\bibitem[{{Andrews} {et~al.}(2016){Andrews}, {Wilner}, {Zhu}, {Birnstiel},
  {Carpenter}, {P{\'e}rez}, {Bai}, {{\"O}berg}, {Hughes}, {Isella}, \&
  {Ricci}}]{andrews2016}
{Andrews}, S.~M., {Wilner}, D.~J., {Zhu}, Z., {et~al.} 2016, \apjl, 820, L40

\bibitem[{{Apai} {et~al.}(2004){Apai}, {Pascucci}, {Brandner}, {Henning},
  {Lenzen}, {Potter}, {Lagrange}, \& {Rousset}}]{apai2004}
{Apai}, D., {Pascucci}, I., {Brandner}, W., {et~al.} 2004, \aap, 415, 671

\bibitem[{{Ataiee} {et~al.}(2013){Ataiee}, {Pinilla}, {Zsom}, {Dullemond},
  {Dominik}, \& {Ghanbari}}]{ataiee2013}
{Ataiee}, S., {Pinilla}, P., {Zsom}, A., {et~al.} 2013, \aap, 553, L3

\bibitem[{{Avenhaus} {et~al.}(2014){Avenhaus}, {Quanz}, {Schmid}, {Meyer},
  {Garufi}, {Wolf}, \& {Dominik}}]{avenhaus2014}
{Avenhaus}, H., {Quanz}, S.~P., {Schmid}, H.~M., {et~al.} 2014, \apj, 781, 87

\bibitem[{{Benisty} {et~al.}(2015){Benisty}, {Juhasz}, {Boccaletti},
  {Avenhaus}, {Milli}, {Thalmann}, {Dominik}, {Pinilla}, {Buenzli}, {Pohl},
  {Beuzit}, {Birnstiel}, {de Boer}, {Bonnefoy}, {Chauvin}, {Christiaens},
  {Garufi}, {Grady}, {Henning}, {Huelamo}, {Isella}, {Langlois}, {M{\'e}nard},
  {Mouillet}, {Olofsson}, {Pantin}, {Pinte}, \& {Pueyo}}]{benisty2015}
{Benisty}, M., {Juhasz}, A., {Boccaletti}, A., {et~al.} 2015, \aap, 578, L6

\bibitem[{{Benisty} {et~al.}(2017){Benisty}, {Stolker}, {Pohl}, {de Boer},
  {Lesur}, {Dominik}, {Dullemond}, {Langlois}, {Min}, {Wagner}, {Henning},
  {Juhasz}, {Pinilla}, {Facchini}, {Apai}, {van Boekel}, {Garufi}, {Ginski},
  {M{\'e}nard}, {Pinte}, {Quanz}, {Zurlo}, {Boccaletti}, {Bonnefoy}, {Beuzit},
  {Chauvin}, {Cudel}, {Desidera}, {Feldt}, {Fontanive}, {Gratton}, {Kasper},
  {Lagrange}, {LeCoroller}, {Mouillet}, {Mesa}, {Sissa}, {Vigan}, {Antichi},
  {Buey}, {Fusco}, {Gisler}, {Llored}, {Magnard}, {Moeller-Nilsson}, {Pragt},
  {Roelfsema}, {Sauvage}, \& {Wildi}}]{benisty2017}
{Benisty}, M., {Stolker}, T., {Pohl}, A., {et~al.} 2017, \aap, 597, A42

\bibitem[{{B{\'e}thune} {et~al.}(2016){B{\'e}thune}, {Lesur}, \&
  {Ferreira}}]{bethune2016}
{B{\'e}thune}, W., {Lesur}, G., \& {Ferreira}, J. 2016, \aap, 589, A87

\bibitem[{{Beuzit} {et~al.}(2008){Beuzit}, {Feldt}, {Dohlen}, {Mouillet},
  {Puget}, {Wildi}, {Abe}, {Antichi}, {Baruffolo}, {Baudoz}, {Boccaletti},
  {Carbillet}, {Charton}, {Claudi}, {Downing}, {Fabron}, {Feautrier},
  {Fedrigo}, {Fusco}, {Gach}, {Gratton}, {Henning}, {Hubin}, {Joos}, {Kasper},
  {Langlois}, {Lenzen}, {Moutou}, {Pavlov}, {Petit}, {Pragt}, {Rabou}, {Rigal},
  {Roelfsema}, {Rousset}, {Saisse}, {Schmid}, {Stadler}, {Thalmann}, {Turatto},
  {Udry}, {Vakili}, \& {Waters}}]{beuzit2008}
{Beuzit}, J.-L., {Feldt}, M., {Dohlen}, K., {et~al.} 2008, in SPIE Proc., Vol.
  7014, 18

\bibitem[{{Biller} {et~al.}(2014){Biller}, {Males}, {Rodigas}, {Morzinski},
  {Close}, {Juh{\'a}sz}, {Follette}, {Lacour}, {Benisty}, {Sicilia-Aguilar},
  {Hinz}, {Weinberger}, {Henning}, {Pott}, {Bonnefoy}, \&
  {K{\"o}hler}}]{biller2014}
{Biller}, B.~A., {Males}, J., {Rodigas}, T., {et~al.} 2014, \apjl, 792, L22

\bibitem[{{Birnstiel} \& {Andrews}(2014)}]{birnstiel2014}
{Birnstiel}, T., \& {Andrews}, S.~M. 2014, \apj, 780, 153

\bibitem[{{Birnstiel} {et~al.}(2010){Birnstiel}, {Dullemond}, \&
  {Brauer}}]{birnstiel2010}
{Birnstiel}, T., {Dullemond}, C.~P., \& {Brauer}, F. 2010, \aap, 513, A79

\bibitem[{{Blondel} \& {Djie}(2006)}]{blondel2006}
{Blondel}, P.~F.~C., \& {Djie}, H.~R.~E.~T.~A. 2006, \aap, 456, 1045

\bibitem[{{Boccaletti} {et~al.}(2008){Boccaletti}, {Abe}, {Baudrand}, {Daban},
  {Douet}, {Guerri}, {Robbe-Dubois}, {Bendjoya}, {Dohlen}, \&
  {Mawet}}]{boccaletti2008}
{Boccaletti}, A., {Abe}, L., {Baudrand}, J., {et~al.} 2008, in \procspie, Vol.
  7015, Adaptive Optics Systems, 70151B

\bibitem[{{Brauer} {et~al.}(2008){Brauer}, {Dullemond}, \&
  {Henning}}]{brauer2008}
{Brauer}, F., {Dullemond}, C.~P., \& {Henning}, T. 2008, \aap, 480, 859

\bibitem[{{Canovas} {et~al.}(2015){Canovas}, {M{\'e}nard}, {de Boer}, {Pinte},
  {Avenhaus}, \& {Schreiber}}]{canovas2015}
{Canovas}, H., {M{\'e}nard}, F., {de Boer}, J., {et~al.} 2015, \aap, 582, L7

\bibitem[{{Canovas} {et~al.}(2017){Canovas}, {Hardy}, {Zurlo}, {Wahhaj},
  {Schreiber}, {Vigan}, {Villaver}, {Olofsson}, {Meeus}, {M{\'e}nard},
  {Caceres}, {Cieza}, \& {Garufi}}]{canovas2017}
{Canovas}, H., {Hardy}, A., {Zurlo}, A., {et~al.} 2017, \aap, 598, A43

\bibitem[{{Carrasco-Gonz{\'a}lez} {et~al.}(2016){Carrasco-Gonz{\'a}lez},
  {Henning}, {Chandler}, {Linz}, {P{\'e}rez}, {Rodr{\'{\i}}guez},
  {Galv{\'a}n-Madrid}, {Anglada}, {Birnstiel}, {van Boekel}, {Flock}, {Klahr},
  {Macias}, {Menten}, {Osorio}, {Testi}, {Torrelles}, \& {Zhu}}]{carrasco2016}
{Carrasco-Gonz{\'a}lez}, C., {Henning}, T., {Chandler}, C.~J., {et~al.} 2016,
  \apjl, 821, L16

\bibitem[{{Christiaens} {et~al.}(2014){Christiaens}, {Casassus}, {Perez}, {van
  der Plas}, \& {M{\'e}nard}}]{christiaens2014}
{Christiaens}, V., {Casassus}, S., {Perez}, S., {van der Plas}, G., \&
  {M{\'e}nard}, F. 2014, \apjl, 785, L12

\bibitem[{{Crida} {et~al.}(2006){Crida}, {Morbidelli}, \& {Masset}}]{crida2006}
{Crida}, A., {Morbidelli}, A., \& {Masset}, F. 2006, \icarus, 181, 587

\bibitem[{{de Boer} {et~al.}(2016){de Boer}, {Salter}, {Benisty}, {Vigan},
  {Boccaletti}, {Pinilla}, {Ginski}, {Juhasz}, {Maire}, {Messina}, {Desidera},
  {Cheetham}, {Girard}, {Wahhaj}, {Langlois}, {Bonnefoy}, {Beuzit}, {Buenzli},
  {Chauvin}, {Dominik}, {Feldt}, {Gratton}, {Hagelberg}, {Isella}, {Janson},
  {Keller}, {Lagrange}, {Lannier}, {Menard}, {Mesa}, {Mouillet}, {Mugrauer},
  {Peretti}, {Perrot}, {Sissa}, {Snik}, {Vogt}, {Zurlo}, \& {SPHERE
  Consortium}}]{deboer2016}
{de Boer}, J., {Salter}, G., {Benisty}, M., {et~al.} 2016, \aap, 595, A114

\bibitem[{{de Juan Ovelar} {et~al.}(2016){de Juan Ovelar}, {Pinilla}, {Min},
  {Dominik}, \& {Birnstiel}}]{dejuanovelar2016}
{de Juan Ovelar}, M., {Pinilla}, P., {Min}, M., {Dominik}, C., \& {Birnstiel},
  T. 2016, \mnras, 459, L85

\bibitem[{{Dipierro} \& {Laibe}(2017)}]{dipierro2017}
{Dipierro}, G., \& {Laibe}, G. 2017, \mnras, 469, 1932

\bibitem[{{Dipierro} {et~al.}(2016){Dipierro}, {Laibe}, {Price}, \&
  {Lodato}}]{dipierro2016}
{Dipierro}, G., {Laibe}, G., {Price}, D.~J., \& {Lodato}, G. 2016, \mnras, 459,
  L1

\bibitem[{{Dodson-Robinson} \& {Salyk}(2011)}]{dodson2011}
{Dodson-Robinson}, S.~E., \& {Salyk}, C. 2011, \apj, 738, 131

\bibitem[{{Dohlen} {et~al.}(2008){Dohlen}, {Langlois}, {Saisse}, {Hill},
  {Origne}, {Jacquet}, {Fabron}, {Blanc}, {Llored}, {Carle}, {Moutou}, {Vigan},
  {Boccaletti}, {Carbillet}, {Mouillet}, \& {Beuzit}}]{dohlen2008}
{Dohlen}, K., {Langlois}, M., {Saisse}, M., {et~al.} 2008, in SPIE Proc., Vol.
  7014

\bibitem[{{Dong} \& {Fung}(2017)}]{dong2017}
{Dong}, R., \& {Fung}, J. 2017, \apj, 835, 146

\bibitem[{{Dong} {et~al.}(2016){Dong}, {Fung}, \& {Chiang}}]{dong2016}
{Dong}, R., {Fung}, J., \& {Chiang}, E. 2016, \apj, 826, 75

\bibitem[{{Dong} {et~al.}(2015){Dong}, {Zhu}, \& {Whitney}}]{dong2015b}
{Dong}, R., {Zhu}, Z., \& {Whitney}, B. 2015, \apj, 809, 93

\bibitem[{{Draine}(2003)}]{draine2003}
{Draine}, B.~T. 2003, \apj, 598, 1026

\bibitem[{{Dullemond} {et~al.}(2012){Dullemond}, {Juhasz}, {Pohl}, {Sereshti},
  {Shetty}, {Peters}, {Commercon}, \& {Flock}}]{radmc}
{Dullemond}, C.~P., {Juhasz}, A., {Pohl}, A., {et~al.} 2012, {RADMC-3D: A
  multi-purpose radiative transfer tool}, Astrophysics Source Code Library,
  ascl:1202.015

\bibitem[{{Dunkin} {et~al.}(1997){Dunkin}, {Barlow}, \& {Ryan}}]{dunkin1997}
{Dunkin}, S.~K., {Barlow}, M.~J., \& {Ryan}, S.~G. 1997, \mnras, 286, 604

\bibitem[{{Dzyurkevich} {et~al.}(2013){Dzyurkevich}, {Turner}, {Henning}, \&
  {Kley}}]{dzyurkevich2013}
{Dzyurkevich}, N., {Turner}, N.~J., {Henning}, T., \& {Kley}, W. 2013, \apj,
  765, 114

\bibitem[{{Facchini} {et~al.}(2017){Facchini}, {Birnstiel}, {Bruderer}, \& {van
  Dishoeck}}]{facchini2017}
{Facchini}, S., {Birnstiel}, T., {Bruderer}, S., \& {van Dishoeck}, E.~F. 2017,
  \aap, 605, A16

\bibitem[{{Fedele} {et~al.}(2017){Fedele}, {Carney}, {Hogerheijde}, {Walsh},
  {Miotello}, {Klaassen}, {Bruderer}, {Henning}, \& {van
  Dishoeck}}]{fedele2017}
{Fedele}, D., {Carney}, M., {Hogerheijde}, M.~R., {et~al.} 2017, \aap, 600, A72

\bibitem[{{Flock} {et~al.}(2015){Flock}, {Ruge}, {Dzyurkevich}, {Henning},
  {Klahr}, \& {Wolf}}]{flock2015}
{Flock}, M., {Ruge}, J.~P., {Dzyurkevich}, N., {et~al.} 2015, \aap, 574, A68

\bibitem[{{Folsom} {et~al.}(2012){Folsom}, {Bagnulo}, {Wade}, {Alecian},
  {Landstreet}, {Marsden}, \& {Waite}}]{folsom2012}
{Folsom}, C.~P., {Bagnulo}, S., {Wade}, G.~A., {et~al.} 2012, \mnras, 422, 2072

\bibitem[{{Fung} {et~al.}(2014){Fung}, {Shi}, \& {Chiang}}]{fung2014}
{Fung}, J., {Shi}, J.-M., \& {Chiang}, E. 2014, \apj, 782, 88

\bibitem[{Fusco {et~al.}(2006)Fusco, Rousset, Sauvage, Petit, Beuzit, Dohlen,
  Mouillet, Charton, Nicolle, Kasper, Baudoz, \& Puget}]{fusco2006}
Fusco, T., Rousset, G., Sauvage, J.-F., {et~al.} 2006, Opt. Express, 14, 7515

\bibitem[{{Gaia Collaboration} {et~al.}(2016){Gaia Collaboration}, {Brown},
  {Vallenari}, {Prusti}, {de Bruijne}, {Mignard}, {Drimmel}, {Babusiaux},
  {Bailer-Jones}, {Bastian}, \& et~al.}]{gaia}
{Gaia Collaboration}, {Brown}, A.~G.~A., {Vallenari}, A., {et~al.} 2016, \aap,
  595, A2

\bibitem[{{Garufi} {et~al.}(2013){Garufi}, {Quanz}, {Avenhaus}, {Buenzli},
  {Dominik}, {Meru}, {Meyer}, {Pinilla}, {Schmid}, \& {Wolf}}]{garufi2013}
{Garufi}, A., {Quanz}, S.~P., {Avenhaus}, H., {et~al.} 2013, \aap, 560, A105

\bibitem[{{Garufi} {et~al.}(2017){Garufi}, {Meeus}, {Benisty}, {Quanz},
  {Banzatti}, {Kama}, {Canovas}, {Eiroa}, {Schmid}, {Stolker}, {Pohl},
  {Rigliaco}, {M{\'e}nard}, {Meyer}, {van Boekel}, \& {Dominik}}]{garufi2017}
{Garufi}, A., {Meeus}, G., {Benisty}, M., {et~al.} 2017, \aap, 603, A21

\bibitem[{{Ginski} {et~al.}(2016){Ginski}, {Stolker}, {Pinilla}, {Dominik},
  {Boccaletti}, {de Boer}, {Benisty}, {Biller}, {Feldt}, {Garufi}, {Keller},
  {Kenworthy}, {Maire}, {M{\'e}nard}, {Mesa}, {Milli}, {Min}, {Pinte}, {Quanz},
  {van Boekel}, {Bonnefoy}, {Chauvin}, {Desidera}, {Gratton}, {Girard},
  {Keppler}, {Kopytova}, {Lagrange}, {Langlois}, {Rouan}, \&
  {Vigan}}]{ginski2016}
{Ginski}, C., {Stolker}, T., {Pinilla}, P., {et~al.} 2016, \aap, 595, A112

\bibitem[{{Grady} {et~al.}(2007){Grady}, {Schneider}, {Hamaguchi}, {Sitko},
  {Carpenter}, {Hines}, {Collins}, {Williger}, {Woodgate}, {Henning},
  {M{\'e}nard}, {Wilner}, {Petre}, {Palunas}, {Quirrenbach}, {Nuth},
  {Silverstone}, \& {Kim}}]{grady2007}
{Grady}, C.~A., {Schneider}, G., {Hamaguchi}, K., {et~al.} 2007, \apj, 665,
  1391

\bibitem[{{Grady} {et~al.}(2013){Grady}, {Muto}, {Hashimoto}, {Fukagawa},
  {Currie}, {Biller}, {Thalmann}, {Sitko}, {Russell}, {Wisniewski}, {Dong},
  {Kwon}, {Sai}, {Hornbeck}, {Schneider}, {Hines}, {Moro Mart{\'{\i}}n},
  {Feldt}, {Henning}, {Pott}, {Bonnefoy}, {Bouwman}, {Lacour}, {Mueller},
  {Juh{\'a}sz}, {Crida}, {Chauvin}, {Andrews}, {Wilner}, {Kraus}, {Dahm},
  {Robitaille}, {Jang-Condell}, {Abe}, {Akiyama}, {Brandner}, {Brandt},
  {Carson}, {Egner}, {Follette}, {Goto}, {Guyon}, {Hayano}, {Hayashi},
  {Hayashi}, {Hodapp}, {Ishii}, {Iye}, {Janson}, {Kandori}, {Knapp}, {Kudo},
  {Kusakabe}, {Kuzuhara}, {Mayama}, {McElwain}, {Matsuo}, {Miyama}, {Morino},
  {Nishimura}, {Pyo}, {Serabyn}, {Suto}, {Suzuki}, {Takami}, {Takato},
  {Terada}, {Tomono}, {Turner}, {Watanabe}, {Yamada}, {Takami}, {Usuda}, \&
  {Tamura}}]{grady2013}
{Grady}, C.~A., {Muto}, T., {Hashimoto}, J., {et~al.} 2013, \apj, 762, 48

\bibitem[{{Kama} {et~al.}(2016){Kama}, {Bruderer}, {Carney}, {Hogerheijde},
  {van Dishoeck}, {Fedele}, {Baryshev}, {Boland}, {G{\"u}sten}, {Aikutalp},
  {Choi}, {Endo}, {Frieswijk}, {Karska}, {Klaassen}, {Koumpia}, {Kristensen},
  {Leurini}, {Nagy}, {Perez Beaupuits}, {Risacher}, {van der Marel}, {van
  Kempen}, {van Weeren}, {Wyrowski}, \& {Y{\i}ld{\i}z}}]{kama2016}
{Kama}, M., {Bruderer}, S., {Carney}, M., {et~al.} 2016, \aap, 588, A108

\bibitem[{{Kanagawa} {et~al.}(2015){Kanagawa}, {Muto}, {Tanaka}, {Tanigawa},
  {Takeuchi}, {Tsukagoshi}, \& {Momose}}]{kanagawa2015}
{Kanagawa}, K.~D., {Muto}, T., {Tanaka}, H., {et~al.} 2015, \apjl, 806, L15

\bibitem[{{Kataoka} {et~al.}(2015){Kataoka}, {Muto}, {Momose}, {Tsukagoshi},
  {Fukagawa}, {Shibai}, {Hanawa}, {Murakawa}, \& {Dullemond}}]{kataoka2015}
{Kataoka}, A., {Muto}, T., {Momose}, M., {et~al.} 2015, \apj, 809, 78

\bibitem[{{Kataoka} {et~al.}(2016){Kataoka}, {Tsukagoshi}, {Momose}, {Nagai},
  {Muto}, {Dullemond}, {Pohl}, {Fukagawa}, {Shibai}, {Hanawa}, \&
  {Murakawa}}]{kataoka2016}
{Kataoka}, A., {Tsukagoshi}, T., {Momose}, M., {et~al.} 2016, \apjl, 831, L12

\bibitem[{{Kuhn} {et~al.}(2001){Kuhn}, {Potter}, \& {Parise}}]{kuhn2001}
{Kuhn}, J.~R., {Potter}, D., \& {Parise}, B. 2001, \apjl, 553, L189

\bibitem[{{Langlois} {et~al.}(2014){Langlois}, {Dohlen}, {Vigan}, {Zurlo},
  {Moutou}, {Schmid}, {Mili}, {Beuzit}, {Boccaletti}, {Carle}, {Costille},
  {Dorn}, {Gluck}, {Hubin}, {Feldt}, {Kasper}, {Lizon}, {Madec}, {Le Mignant},
  {Mouillet}, {Puget}, {Sauvage}, \& {Wildi}}]{langlois2014}
{Langlois}, M., {Dohlen}, K., {Vigan}, A., {et~al.} 2014, in SPIE Proc., Vol.
  9147, 1

\bibitem[{{Lazareff} {et~al.}(2017){Lazareff}, {Berger}, {Kluska}, {Le
  Bouquin}, {Benisty}, {Malbet}, {Koen}, {Pinte}, {Thi}, {Absil}, {Baron},
  {Delboulb{\'e}}, {Duvert}, {Isella}, {Jocou}, {Juhasz}, {Kraus}, {Lachaume},
  {M{\'e}nard}, {Millan-Gabet}, {Monnier}, {Moulin}, {Perraut}, {Rochat},
  {Soulez}, {Tallon}, {Thi{\'e}baut}, {Traub}, \& {Zins}}]{lazareff2017}
{Lazareff}, B., {Berger}, J.-P., {Kluska}, J., {et~al.} 2017, \aap, 599, A85

\bibitem[{{Ligi} {et~al.}(2017){Ligi}, {Vigan}, {Gratton}, {de Boer},
  {Benisty}, {Boccaletti}, {Quanz}, {Meyer}, {Ginski}, {Sissa}, {Henning},
  {Beuzit}, {Biller}, {Bonnefoy}, {Chauvin}, {Cheetham}, {Cudel}, {Delorme},
  {Desidera}, {Feldt}, {Galicher}, {Girard}, {Janson}, {Kasper}, {Kopytova},
  {Lagrange}, {Langlois}, {Lecoroller}, {Maire}, {M{\'e}nard}, {Mesa},
  {Peretti}, {Perrot}, {Pinilla}, {Pohl}, {Rouan}, {Stolker}, {Samland},
  {Wahhaj}, {Wildi}, {Zurlo}, {Buey}, {Fantinel}, {Fusco}, {Jaquet}, {Moulin},
  {Ramos}, {Suarez}, \& {Weber}}]{ligi2017}
{Ligi}, R., {Vigan}, A., {Gratton}, R., {et~al.} 2017, ArXiv e-prints,
  arXiv:1709.01734

\bibitem[{{Mac{\'{\i}}as} {et~al.}(2017){Mac{\'{\i}}as}, {Anglada}, {Osorio},
  {Torrelles}, {Carrasco-Gonz{\'a}lez}, {G{\'o}mez}, {Rodr{\'{\i}}guez}, \&
  {Sierra}}]{macias2017}
{Mac{\'{\i}}as}, E., {Anglada}, G., {Osorio}, M., {et~al.} 2017, \apj, 838, 97

\bibitem[{{Maire} {et~al.}(2016){Maire}, {Langlois}, {Dohlen}, {Lagrange},
  {Gratton}, {Chauvin}, {Desidera}, {Girard}, {Milli}, {Vigan}, {Zins},
  {Delorme}, {Beuzit}, {Claudi}, {Feldt}, {Mouillet}, {Puget}, {Turatto}, \&
  {Wildi}}]{maire2016}
{Maire}, A.-L., {Langlois}, M., {Dohlen}, K., {et~al.} 2016, in \procspie, Vol.
  9908, Ground-based and Airborne Instrumentation for Astronomy VI, 990834

\bibitem[{{Marino} {et~al.}(2015){Marino}, {Perez}, \& {Casassus}}]{marino2015}
{Marino}, S., {Perez}, S., \& {Casassus}, S. 2015, \apjl, 798, L44

\bibitem[{{Meeus} {et~al.}(2010){Meeus}, {Pinte}, {Woitke}, {Montesinos},
  {Mendigut{\'{\i}}a}, {Riviere-Marichalar}, {Eiroa}, {Mathews},
  {Vandenbussche}, {Howard}, {Roberge}, {Sandell}, {Duch{\^e}ne}, {M{\'e}nard},
  {Grady}, {Dent}, {Kamp}, {Augereau}, {Thi}, {Tilling}, {Alacid}, {Andrews},
  {Ardila}, {Aresu}, {Barrado}, {Brittain}, {Ciardi}, {Danchi}, {Fedele}, {de
  Gregorio-Monsalvo}, {Heras}, {Huelamo}, {Krivov}, {Lebreton}, {Liseau},
  {Martin-Zaidi}, {Mora}, {Morales-Calderon}, {Nomura}, {Pantin}, {Pascucci},
  {Phillips}, {Podio}, {Poelman}, {Ramsay}, {Riaz}, {Rice}, {Solano}, {Walker},
  {White}, {Williams}, \& {Wright}}]{meeus2010}
{Meeus}, G., {Pinte}, C., {Woitke}, P., {et~al.} 2010, \aap, 518, L124

\bibitem[{{Menu} {et~al.}(2015){Menu}, {van Boekel}, {Henning}, {Leinert},
  {Waelkens}, \& {Waters}}]{menu2015}
{Menu}, J., {van Boekel}, R., {Henning}, T., {et~al.} 2015, \aap, 581, A107

\bibitem[{{Momose} {et~al.}(2015){Momose}, {Morita}, {Fukagawa}, {Muto},
  {Takeuchi}, {Hashimoto}, {Honda}, {Kudo}, {Okamoto}, {Kanagawa}, {Tanaka},
  {Grady}, {Sitko}, {Akiyama}, {Currie}, {Follette}, {Mayama}, {Kusakabe},
  {Abe}, {Brandner}, {Brandt}, {Carson}, {Egner}, {Feldt}, {Goto}, {Guyon},
  {Hayano}, {Hayashi}, {Hayashi}, {Henning}, {Hodapp}, {Ishii}, {Iye},
  {Janson}, {Kandori}, {Knapp}, {Kuzuhara}, {Kwon}, {Matsuo}, {McElwain},
  {Miyama}, {Morino}, {Moro-Martin}, {Nishimura}, {Pyo}, {Serabyn}, {Suenaga},
  {Suto}, {Suzuki}, {Takahashi}, {Takami}, {Takato}, {Terada}, {Thalmann},
  {Tomono}, {Turner}, {Watanabe}, {Wisniewski}, {Yamada}, {Takami}, {Usuda}, \&
  {Tamura}}]{momose2015}
{Momose}, M., {Morita}, A., {Fukagawa}, M., {et~al.} 2015, \pasj, 67, 83

\bibitem[{{Monnier} {et~al.}(2017){Monnier}, {Harries}, {Aarnio}, {Adams},
  {Andrews}, {Calvet}, {Espaillat}, {Hartmann}, {Hinkley}, {Kraus}, {McClure},
  {Oppenheimer}, {Perrin}, \& {Wilner}}]{monnier2017}
{Monnier}, J.~D., {Harries}, T.~J., {Aarnio}, A., {et~al.} 2017, \apj, 838, 20

\bibitem[{{Mordasini} {et~al.}(2012){Mordasini}, {Alibert}, {Klahr}, \&
  {Henning}}]{mordasini2012}
{Mordasini}, C., {Alibert}, Y., {Klahr}, H., \& {Henning}, T. 2012, \aap, 547,
  A111

\bibitem[{{Mordasini} {et~al.}(2016){Mordasini}, {van Boekel}, {Molli{\`e}re},
  {Henning}, \& {Benneke}}]{mordasini2016}
{Mordasini}, C., {van Boekel}, R., {Molli{\`e}re}, P., {Henning}, T., \&
  {Benneke}, B. 2016, \apj, 832, 41

\bibitem[{{Muto} {et~al.}(2012){Muto}, {Grady}, {Hashimoto}, {Fukagawa},
  {Hornbeck}, {Sitko}, {Russell}, {Werren}, {Cur{\'e}}, {Currie}, {Ohashi},
  {Okamoto}, {Momose}, {Honda}, {Inutsuka}, {Takeuchi}, {Dong}, {Abe},
  {Brandner}, {Brandt}, {Carson}, {Egner}, {Feldt}, {Fukue}, {Goto}, {Guyon},
  {Hayano}, {Hayashi}, {Hayashi}, {Henning}, {Hodapp}, {Ishii}, {Iye},
  {Janson}, {Kandori}, {Knapp}, {Kudo}, {Kusakabe}, {Kuzuhara}, {Matsuo},
  {Mayama}, {McElwain}, {Miyama}, {Morino}, {Moro-Martin}, {Nishimura}, {Pyo},
  {Serabyn}, {Suto}, {Suzuki}, {Takami}, {Takato}, {Terada}, {Thalmann},
  {Tomono}, {Turner}, {Watanabe}, {Wisniewski}, {Yamada}, {Takami}, {Usuda}, \&
  {Tamura}}]{muto2012}
{Muto}, T., {Grady}, C.~A., {Hashimoto}, J., {et~al.} 2012, \apjl, 748, L22

\bibitem[{{Okuzumi} {et~al.}(2016){Okuzumi}, {Momose}, {Sirono}, {Kobayashi},
  \& {Tanaka}}]{okuzumi2016}
{Okuzumi}, S., {Momose}, M., {Sirono}, S.-i., {Kobayashi}, H., \& {Tanaka}, H.
  2016, \apj, 821, 82

\bibitem[{{Osorio} {et~al.}(2014){Osorio}, {Anglada}, {Carrasco-Gonz{\'a}lez},
  {Torrelles}, {Mac{\'{\i}}as}, {Rodr{\'{\i}}guez}, {G{\'o}mez}, {D'Alessio},
  {Calvet}, {Nagel}, {Dent}, {Quanz}, {Reggiani}, \&
  {Mayen-Gijon}}]{osorio2014}
{Osorio}, M., {Anglada}, G., {Carrasco-Gonz{\'a}lez}, C., {et~al.} 2014, \apjl,
  791, L36

\bibitem[{{Paardekooper} \& {Mellema}(2004)}]{paardekooper2004}
{Paardekooper}, S.-J., \& {Mellema}, G. 2004, \aap, 425, L9

\bibitem[{{Paardekooper} \& {Mellema}(2006)}]{paardekooper2006}
---. 2006, \aap, 453, 1129

\bibitem[{{Pani{\'c}} {et~al.}(2008){Pani{\'c}}, {Hogerheijde}, {Wilner}, \&
  {Qi}}]{panic2008}
{Pani{\'c}}, O., {Hogerheijde}, M.~R., {Wilner}, D., \& {Qi}, C. 2008, \aap,
  491, 219

\bibitem[{{P{\'e}rez} {et~al.}(2016){P{\'e}rez}, {Carpenter}, {Andrews},
  {Ricci}, {Isella}, {Linz}, {Sargent}, {Wilner}, {Henning}, {Deller},
  {Chandler}, {Dullemond}, {Lazio}, {Menten}, {Corder}, {Storm}, {Testi},
  {Tazzari}, {Kwon}, {Calvet}, {Greaves}, {Harris}, \& {Mundy}}]{perez2016}
{P{\'e}rez}, L.~M., {Carpenter}, J.~M., {Andrews}, S.~M., {et~al.} 2016,
  Science, 353, 1519

\bibitem[{{Petit} {et~al.}(2014){Petit}, {Sauvage}, {Fusco}, {Sevin}, {Suarez},
  {Costille}, {Vigan}, {Soenke}, {Perret}, {Rochat}, {Barrufolo}, {Salasnich},
  {Beuzit}, {Dohlen}, {Mouillet}, {Puget}, {Wildi}, {Kasper}, {Conan},
  {Kulcs{\'a}r}, \& {Raynaud}}]{petit2014}
{Petit}, C., {Sauvage}, J.-F., {Fusco}, T., {et~al.} 2014, in SPIE Proc., Vol.
  9148, 0

\bibitem[{{Picogna} \& {Kley}(2015)}]{picogna2015}
{Picogna}, G., \& {Kley}, W. 2015, \aap, 584, A110

\bibitem[{{Pinilla} {et~al.}(2012){Pinilla}, {Benisty}, \&
  {Birnstiel}}]{pinilla2012}
{Pinilla}, P., {Benisty}, M., \& {Birnstiel}, T. 2012, \aap, 545, A81

\bibitem[{{Pinilla} {et~al.}(2015{\natexlab{a}}){Pinilla}, {Birnstiel}, \&
  {Walsh}}]{pinilla2015b}
{Pinilla}, P., {Birnstiel}, T., \& {Walsh}, C. 2015{\natexlab{a}}, \aap, 580,
  A105

\bibitem[{{Pinilla} {et~al.}(2015{\natexlab{b}}){Pinilla}, {de Juan Ovelar},
  {Ataiee}, {Benisty}, {Birnstiel}, {van Dishoeck}, \& {Min}}]{pinilla2015a}
{Pinilla}, P., {de Juan Ovelar}, M., {Ataiee}, S., {et~al.} 2015{\natexlab{b}},
  \aap, 573, A9

\bibitem[{{Pinilla} {et~al.}(2016){Pinilla}, {Flock}, {Ovelar}, \&
  {Birnstiel}}]{pinilla2016}
{Pinilla}, P., {Flock}, M., {Ovelar}, M.~d.~J., \& {Birnstiel}, T. 2016, \aap,
  596, A81

\bibitem[{{Pinilla} {et~al.}(2017){Pinilla}, {Pohl}, {Stammler}, \&
  {Birnstiel}}]{pinilla2017}
{Pinilla}, P., {Pohl}, A., {Stammler}, S.~M., \& {Birnstiel}, T. 2017, \apj,
  845, 68

\bibitem[{{Pohl} {et~al.}(2016){Pohl}, {Kataoka}, {Pinilla}, {Dullemond},
  {Henning}, \& {Birnstiel}}]{pohl2016}
{Pohl}, A., {Kataoka}, A., {Pinilla}, P., {et~al.} 2016, \aap, 593, A12

\bibitem[{{Pohl} {et~al.}(2017){Pohl}, {Sissa}, {Langlois}, {M{\"u}ller},
  {Ginski}, {van Holstein}, {Vigan}, {Mesa}, {Maire}, {Henning}, {Gratton},
  {Olofsson}, {van Boekel}, {Benisty}, {Biller}, {Boccaletti}, {Chauvin},
  {Daemgen}, {de Boer}, {Desidera}, {Dominik}, {Garufi}, {Janson}, {Kral},
  {M{\'e}nard}, {Pinte}, {Stolker}, {Szul{\'a}gyi}, {Zurlo}, {Bonnefoy},
  {Cheetham}, {Cudel}, {Feldt}, {Kasper}, {Lagrange}, {Perrot}, \&
  {Wildi}}]{pohl2017}
{Pohl}, A., {Sissa}, E., {Langlois}, M., {et~al.} 2017, \aap, 605, A34

\bibitem[{{Quanz} {et~al.}(2013){Quanz}, {Avenhaus}, {Buenzli}, {Garufi},
  {Schmid}, \& {Wolf}}]{quanz2013}
{Quanz}, S.~P., {Avenhaus}, H., {Buenzli}, E., {et~al.} 2013, \apjl, 766, L2

\bibitem[{{Raman} {et~al.}(2006){Raman}, {Lisanti}, {Wilner}, {Qi}, \&
  {Hogerheijde}}]{raman2006}
{Raman}, A., {Lisanti}, M., {Wilner}, D.~J., {Qi}, C., \& {Hogerheijde}, M.
  2006, \aj, 131, 2290

\bibitem[{{Rapson} {et~al.}(2015){Rapson}, {Kastner}, {Millar-Blanchaer}, \&
  {Dong}}]{rapson2015}
{Rapson}, V.~A., {Kastner}, J.~H., {Millar-Blanchaer}, M.~A., \& {Dong}, R.
  2015, \apjl, 815, L26

\bibitem[{{Reggiani} {et~al.}(2014){Reggiani}, {Quanz}, {Meyer}, {Pueyo},
  {Absil}, {Amara}, {Anglada}, {Avenhaus}, {Girard}, {Carrasco Gonzalez},
  {Graham}, {Mawet}, {Meru}, {Milli}, {Osorio}, {Wolff}, \&
  {Torrelles}}]{reggiani2014}
{Reggiani}, M., {Quanz}, S.~P., {Meyer}, M.~R., {et~al.} 2014, \apjl, 792, L23

\bibitem[{{Ricci} {et~al.}(2010){Ricci}, {Testi}, {Natta}, {Neri}, {Cabrit}, \&
  {Herczeg}}]{ricci2010}
{Ricci}, L., {Testi}, L., {Natta}, A., {et~al.} 2010, \aap, 512, A15

\bibitem[{{Riviere-Marichalar} {et~al.}(2016){Riviere-Marichalar},
  {Mer{\'{\i}}n}, {Kamp}, {Eiroa}, \& {Montesinos}}]{rivieremarichalar2016}
{Riviere-Marichalar}, P., {Mer{\'{\i}}n}, B., {Kamp}, I., {Eiroa}, C., \&
  {Montesinos}, B. 2016, \aap, 594, A59

\bibitem[{{Rosotti} {et~al.}(2016){Rosotti}, {Juhasz}, {Booth}, \&
  {Clarke}}]{rosotti2016}
{Rosotti}, G.~P., {Juhasz}, A., {Booth}, R.~A., \& {Clarke}, C.~J. 2016,
  \mnras, 459, 2790

\bibitem[{{Ruge} {et~al.}(2016){Ruge}, {Flock}, {Wolf}, {Dzyurkevich},
  {Fromang}, {Henning}, {Klahr}, \& {Meheut}}]{ruge2016}
{Ruge}, J.~P., {Flock}, M., {Wolf}, S., {et~al.} 2016, \aap, 590, A17

\bibitem[{{Sauvage} {et~al.}(2014){Sauvage}, {Fusco}, {Petit}, {Mouillet},
  {Dohlen}, {Costille}, {Beuzit}, {Baruffolo}, {Kasper}, {Suarez Valles},
  {Downing}, {Feautrier}, {Mugnier}, \& {Baudoz}}]{sauvage2014}
{Sauvage}, J., {Fusco}, T., {Petit}, C., {et~al.} 2014, in SPIE Proc., Vol.
  9148

\bibitem[{{Schmid} {et~al.}(2006){Schmid}, {Joos}, \& {Tschan}}]{schmid2006}
{Schmid}, H.~M., {Joos}, F., \& {Tschan}, D. 2006, \aap, 452, 657

\bibitem[{{Schmid} {et~al.}(2012){Schmid}, {Downing}, {Roelfsema}, {Bazzon},
  {Gisler}, {Pragt}, {Cumani}, {Salasnich}, {Pavlov}, {Baruffolo}, {Beuzit},
  {Costille}, {Deiries}, {Dohlen}, {Dominik}, {Elswijk}, {Feldt}, {Kasper},
  {Mouillet}, {Thalmann}, \& {Wildi}}]{schmid2012}
{Schmid}, H.-M., {Downing}, M., {Roelfsema}, R., {et~al.} 2012, in \procspie,
  Vol. 8446, Ground-based and Airborne Instrumentation for Astronomy IV, 84468Y

\bibitem[{{Seok} \& {Li}(2017)}]{seok2017}
{Seok}, J.~Y., \& {Li}, A. 2017, \apj, 835, 291

\bibitem[{{Simon} \& {Armitage}(2014)}]{simon2014}
{Simon}, J.~B., \& {Armitage}, P.~J. 2014, \apj, 784, 15

\bibitem[{{Stammler} {et~al.}(2017){Stammler}, {Birnstiel}, {Pani{\'c}},
  {Dullemond}, \& {Dominik}}]{stammler2017}
{Stammler}, S.~M., {Birnstiel}, T., {Pani{\'c}}, O., {Dullemond}, C.~P., \&
  {Dominik}, C. 2017, \aap, 600, A140

\bibitem[{{Stolker} {et~al.}(2016){Stolker}, {Dominik}, {Avenhaus}, {Min}, {de
  Boer}, {Ginski}, {Schmid}, {Juhasz}, {Bazzon}, {Waters}, {Garufi},
  {Augereau}, {Benisty}, {Boccaletti}, {Henning}, {Langlois}, {Maire},
  {M{\'e}nard}, {Meyer}, {Pinte}, {Quanz}, {Thalmann}, {Beuzit}, {Carbillet},
  {Costille}, {Dohlen}, {Feldt}, {Gisler}, {Mouillet}, {Pavlov}, {Perret},
  {Petit}, {Pragt}, {Rochat}, {Roelfsema}, {Salasnich}, {Soenke}, \&
  {Wildi}}]{stolker2016}
{Stolker}, T., {Dominik}, C., {Avenhaus}, H., {et~al.} 2016, \aap, 595, A113

\bibitem[{{Strom} {et~al.}(1989){Strom}, {Strom}, {Edwards}, {Cabrit}, \&
  {Skrutskie}}]{strom1989}
{Strom}, K.~M., {Strom}, S.~E., {Edwards}, S., {Cabrit}, S., \& {Skrutskie},
  M.~F. 1989, \aj, 97, 1451

\bibitem[{{Takahashi} \& {Inutsuka}(2014)}]{takahashi2014}
{Takahashi}, S.~Z., \& {Inutsuka}, S.-i. 2014, \apj, 794, 55

\bibitem[{{Tang} {et~al.}(2017){Tang}, {Guilloteau}, {Dutrey}, {Muto}, {Shen},
  {Gu}, {Inutsuka}, {Momose}, {Pietu}, {Fukagawa}, {Chapillon}, {Ho}, {di
  Folco}, {Corder}, {Ohashi}, \& {Hashimoto}}]{tang2017}
{Tang}, Y.-W., {Guilloteau}, S., {Dutrey}, A., {et~al.} 2017, \apj, 840, 32

\bibitem[{{Thalmann} {et~al.}(2008){Thalmann}, {Schmid}, {Boccaletti},
  {Mouillet}, {Dohlen}, {Roelfsema}, {Carbillet}, {Gisler}, {Beuzit}, {Feldt},
  {Gratton}, {Joos}, {Keller}, {Kragt}, {Pragt}, {Puget}, {Rigal}, {Snik},
  {Waters}, \& {Wildi}}]{thalmann2008}
{Thalmann}, C., {Schmid}, H.~M., {Boccaletti}, A., {et~al.} 2008, in \procspie,
  Vol. 7014, Ground-based and Airborne Instrumentation for Astronomy II, 70143F

\bibitem[{{Tsukagoshi} {et~al.}(2016){Tsukagoshi}, {Nomura}, {Muto}, {Kawabe},
  {Ishimoto}, {Kanagawa}, {Okuzumi}, {Ida}, {Walsh}, \&
  {Millar}}]{tsukagoshi2016}
{Tsukagoshi}, T., {Nomura}, H., {Muto}, T., {et~al.} 2016, \apjl, 829, L35

\bibitem[{{van Boekel} {et~al.}(2017){van Boekel}, {Henning}, {Menu}, {de
  Boer}, {Langlois}, {M{\"u}ller}, {Avenhaus}, {Boccaletti}, {Schmid},
  {Thalmann}, {Benisty}, {Dominik}, {Ginski}, {Girard}, {Gisler}, {Lobo Gomes},
  {Menard}, {Min}, {Pavlov}, {Pohl}, {Quanz}, {Rabou}, {Roelfsema}, {Sauvage},
  {Teague}, {Wildi}, \& {Zurlo}}]{vanboekel2017}
{van Boekel}, R., {Henning}, T., {Menu}, J., {et~al.} 2017, \apj, 837, 132

\bibitem[{{van der Plas} {et~al.}(2017){van der Plas}, {Wright}, {M{\'e}nard},
  {Casassus}, {Canovas}, {Pinte}, {Maddison}, {Maaskant}, {Avenhaus}, {Cieza},
  {Perez}, \& {Ubach}}]{vanderplas2017}
{van der Plas}, G., {Wright}, C.~M., {M{\'e}nard}, F., {et~al.} 2017, \aap,
  597, A32

\bibitem[{{Wagner} {et~al.}(2015){Wagner}, {Sitko}, {Grady}, {Swearingen},
  {Champney}, {Johnson}, {Werren}, {Whitney}, {Russell}, {Schneider}, {Momose},
  {Muto}, {Inoue}, {Lauroesch}, {Hornbeck}, {Brown}, {Fukagawa}, {Currie},
  {Wisniewski}, \& {Woodgate}}]{wagner2015}
{Wagner}, K.~R., {Sitko}, M.~L., {Grady}, C.~A., {et~al.} 2015, \apj, 798, 94

\bibitem[{{Warren} \& {Brandt}(2008)}]{warren2008}
{Warren}, S.~G., \& {Brandt}, R.~E. 2008, Journal of Geophysical Research
  (Atmospheres), 113, D14220

\bibitem[{{Williams} \& {Cieza}(2011)}]{williams2011}
{Williams}, J.~P., \& {Cieza}, L.~A. 2011, \araa, 49, 67

\bibitem[{{Yang} {et~al.}(2016){Yang}, {Li}, {Looney}, \&
  {Stephens}}]{yang2016}
{Yang}, H., {Li}, Z.-Y., {Looney}, L., \& {Stephens}, I. 2016, \mnras, 456,
  2794

\bibitem[{{Youdin}(2011)}]{youdin2011}
{Youdin}, A.~N. 2011, \apj, 731, 99

\bibitem[{{Zhang} {et~al.}(2015){Zhang}, {Blake}, \& {Bergin}}]{zhang2015}
{Zhang}, K., {Blake}, G.~A., \& {Bergin}, E.~A. 2015, \apjl, 806, L7

\bibitem[{{Zhu} {et~al.}(2012){Zhu}, {Nelson}, {Dong}, {Espaillat}, \&
  {Hartmann}}]{zhu2012}
{Zhu}, Z., {Nelson}, R.~P., {Dong}, R., {Espaillat}, C., \& {Hartmann}, L.
  2012, \apj, 755, 6

\bibitem[{{Zhu} {et~al.}(2011){Zhu}, {Nelson}, {Hartmann}, {Espaillat}, \&
  {Calvet}}]{zhu2011}
{Zhu}, Z., {Nelson}, R.~P., {Hartmann}, L., {Espaillat}, C., \& {Calvet}, N.
  2011, \apj, 729, 47

\bibitem[{{Zubko} {et~al.}(1996){Zubko}, {Mennella}, {Colangeli}, \&
  {Bussoletti}}]{zubko1996}
{Zubko}, V.~G., {Mennella}, V., {Colangeli}, L., \& {Bussoletti}, E. 1996,
  \mnras, 282, 1321

\end{thebibliography}

\end{document}